%% file: main.tex
\documentclass[reprint,
superscriptaddress,
nofootinbib,
 amsmath,amssymb,
 aps,
a4paper,twocolumn, floatfix
]{revtex4-2}
\usepackage{dsfont}
\usepackage{booktabs}
\usepackage{graphicx}
\usepackage{dcolumn}
\usepackage{bm}
\usepackage[hidelinks]{hyperref}
\hypersetup{
    colorlinks,
    linkcolor={red!50!black},
    citecolor={blue!50!black},
    urlcolor={blue!80!black}
}

\usepackage[mathlines]{lineno}

\newcommand{\ket}[1]{|#1\rangle}
\newcommand{\cket}[1]{|#1\rangle_2}
\newcommand{\bra}[1]{\langle #1 |}
\newcommand{\cbra}[1]{\langle #1 |_2}

\newcommand{\shiftleft}[2]{\makebox[0pt][r]{\makebox[#1][l]{#2}}}

\usepackage{tikz}
\usetikzlibrary{matrix,backgrounds}
\pgfdeclarelayer{myback}
\pgfsetlayers{myback,background,main}

\tikzset{mycolor/.style = {line width=1bp,color=#1}}%
\tikzset{myfillcolor/.style = {draw,fill=#1}}%

\NewDocumentCommand{\highlight}{O{blue!40} m m}{%
\draw[mycolor=#1] (#2.north west)rectangle (#3.south east);
}

\newcommand{\Tr}{\text{Tr}}

\newcommand{\fs}[1]{\textcolor{black}{#1}}
\newcommand{\DG}[1]{\textcolor{black}{#1}}

\begin{document}

\preprint{APS/123-QED}

\title{Deterministic entangling gates with nonlinear quantum photonic interferometers}

\author{Francsco Scala}
 \email{francesco.scala01@ateneopv.it}
\author{Davide Nigro}%
\author{Dario Gerace}%
\affiliation{%
Dipartimento di Fisica, Universit\`a degli Studi di Pavia, via Bassi 6, 27100 Pavia (Italy)
}%





\begin{abstract}
The quantum computing paradigm in photonics currently relies on the multi-port interference in linear optical devices, which is intrinsically based on probabilistic measurements outcome and thus non-deterministic. Devising a fully deterministic, universal, and practically achievable quantum computing platform based on integrated photonic circuits is still an open challenge. 
Here we propose to exploit weakly nonlinear photonic devices to implement deterministic entangling quantum gates, following the definition of dual rail photonic qubits. It is shown that a universal set of single- and two-qubit gates can be designed by a suitable concatenation of few optical interferometric elements, with optimal fidelities arbitrarily close to 100\% theoretically demonstrated through a bound constrained optimization algorithm. The actual realization would require the concatenation of a few tens of elementary operations, as well as on-chip optical nonlinearities that are compatible \DG{with some of the existing quantum photonic platforms, as it is finally discussed}. 
\end{abstract}

\maketitle

\section{Introduction}
\label{sec:intro}

Universal quantum computation (QC) can be realized by combining arbitrary single-qubit rotations with two-qubit  entangling gates, such as the controlled-NOT (CNOT) quantum gate. So far, there has been a lot of effort to build the theoretical model for a practical universal photonic quantum computer~\cite{kok_linear_2007,Couteau_2023, Thompson_2011, takeda_toward_2019} harnessing non-classical interference and only employing linear optical elements, but this only allows to obtain a non-deterministic version of the desired entangling gate~\cite{Knill_2001,ralph_linear_2002,okamoto_realization_2011}, i.e., this quantum gate can be implemented with arbitrarily high fidelity but only with probabilistic outcome. 
Even though a deterministic CNOT could be implemented, in principle, by exploiting strongly nonlinear optical elements at the level of single photons,~\cite{milburn_quantum_1989,turchette_measurement_1995,Rauschenbeutel1999coherent, resch2002conditional} there are still considerable difficulties in realizing sufficiently large photon-photon interactions allowing to achieve the so-called ``photon blockade'' regime~\cite{Werner1999,Verger2006,Ferretti2012,Majumdar2013,Gerace-NV-nmat-2019}. Even if promising proof-of-concept demonstrations have been shown in solid state cavity QED~\cite{Faraon2008,Reinhard2012,Delteil2019,Munoz-Matutano2019,Najer2019,Lu2020}, and interesting quantum photonic devices might be realized based on these outcomes~\cite{Chang2007,Gerace2009,Gullans2013}, the actual possibility of exploiting single-photon nonlinearities for developing universal quantum computation has been debated in the past~\cite{Shapiro2006}. Alternative proposals to implement deterministic two-qubit quantum gates have been recently put forward in various nonlinear photonic platforms~\cite{Lahini_2018,Cala19,Englund_PRL_2020,Ghosh2020,Nigro_2022}, but no significant experimental proof-of-principle demonstration has been published, so far. Hence, the question naturally arises if deterministic QC in photonic platforms can be still considered a viable route, e.g., by exploiting basic interferometric elements, as in linear optics quantum computation, combined with weak values of photon nonlinearities. 

Following the routes traced in ~\cite{Nigro_2022}, \DG{where integrated polariton interferometers were proposed to realize deterministic quantum gates using \textit{single-rail} encoding, here we go much further than the previous work by introducing} a novel QC paradigm based on generalized quantum photonic interferometers, which are shown to allow for an efficient implementation of single and two-qubits gates based on \DG{\textit{dual-rail}} qubits encoding. The platform requirements can be reduced to a standard planar technology in which propagating single photons interact {with realistically achievable degree of nonlinearity} when simultaneously present in the same waveguide channel, and freely propagate otherwise. By nontrivially combining a few elementary layers, namely free propagation and hopping regions, we show by numerical optimization that deterministic two-qubit gate fidelities arbitrarily close to 100\% can be achieved, despite the moderate values of photon-photon nonlinear coupling.
We also notice that our approach does not require expanding the Hilbert space of the system by introducing, e.g., ancillary degrees of freedom~\cite{Lahini_2018}.
Interestingly, it is shown that the optimization algorithms achieve the most successful implementation of two-qubit gates for intermediate values of photon-photon nonlinearity, which is  explicitly shown for the paradigmatic cases of CNOT and M\o{}lmer-S\o{}rensen (M-S)~\cite{molmer-sorensen1,molmer-sorensen} entangling gates. These can be combined with arbitrary single-qubit rotations on the Bloch sphere (in particular, $x-$ and $z-$rotations) to implement a universal set of quantum gates, and thus a full QC architecture. 
\DG{Finally, we discuss the practical feasibility of the proposed scheme, by carefully analysing the effects of losses, decoherence, fabrication tolerance of the expected gate fidelity, for which we report extensive results in the Appendix~\ref{appendix:incoherent dyn} and \ref{appendix:tolerance}}.
This result may ultimately open the route to the realization of deterministic quantum photonic computing.

\section{Results}

A general formalism to analyse quantum photonic interferometers in the presence of photon-photon nonlinearities has been previously introduced~\cite{Nigro_2022}. There, the formalism was meant to theoretically describe integrated platforms in which exciton-polaritons are the propagating elementary excitations, owing to their superior nonlinear properties as compared to standard optoelectronics materials. However, the formalism can be generally transferred to any material platform possessing an intrinsic third-order nonlinearity, which may be suitably enhanced by dielectric confinement and being described by a Kerr-type nonlinear Hamiltonian in second quantization. Hence, in the present work we prefer to keep the theoretical discussion as general as possible and speak about ``interacting photons''; we will specifically refer to the potentially targeted platforms in the Discussion section.

\begin{figure}[t]
    \centering
    \includegraphics[trim={0cm 0cm 0cm 0cm},clip,width=0.3\textwidth]{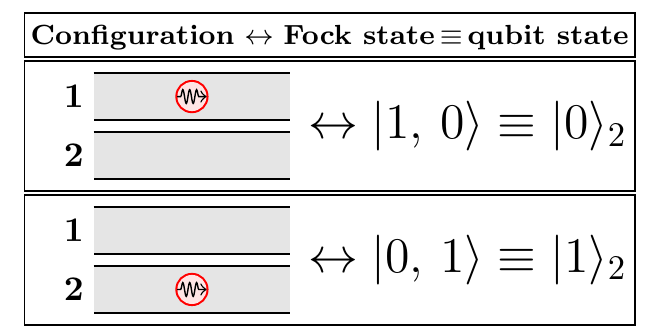}
    \shiftleft{7cm}{\raisebox{2.3cm}[0cm][0cm]{(a)}}
    \includegraphics[trim={0 0 0 0},clip,width=0.4\textwidth]{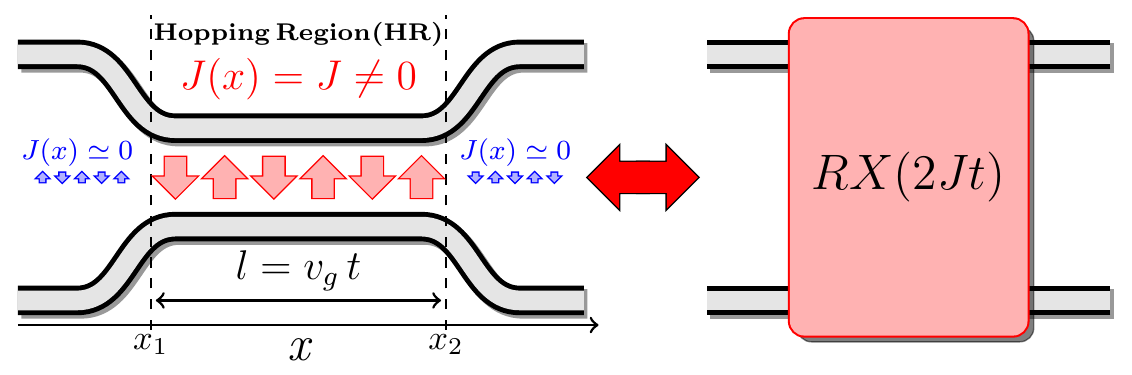}
    \shiftleft{8.cm}{\raisebox{1.8cm}[0cm][0cm]{(b)}}
    \includegraphics[width=0.45\textwidth]{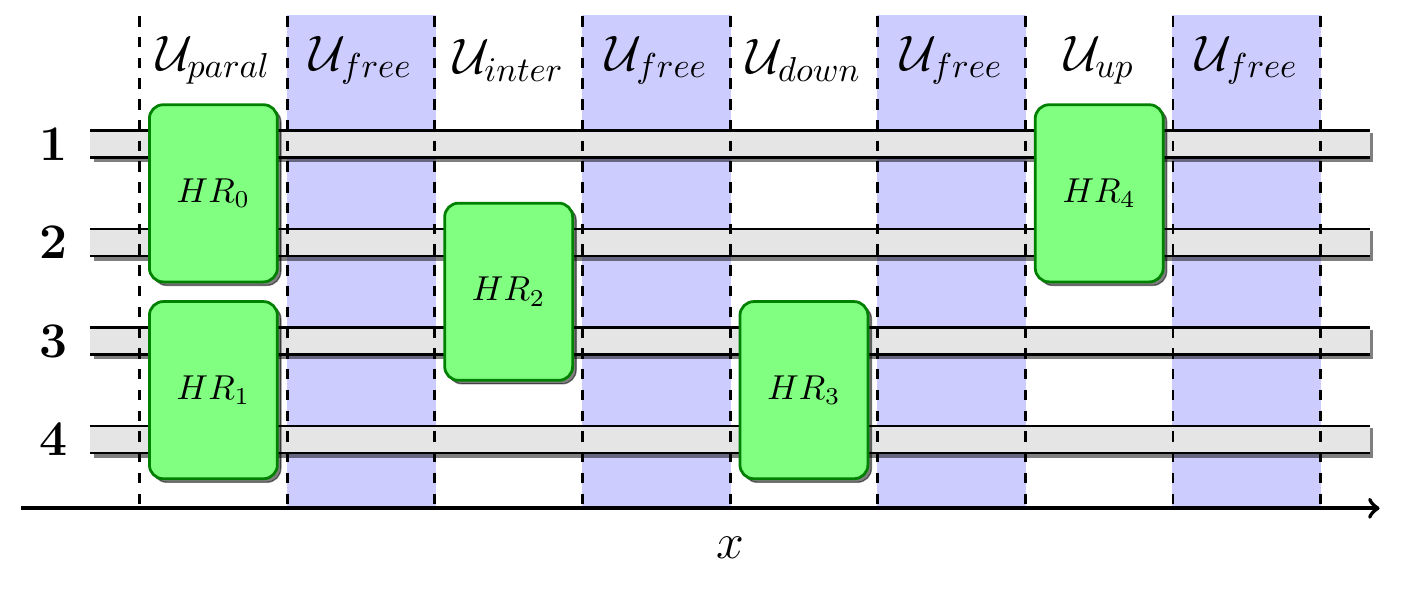}
    \shiftleft{8.45cm}{\raisebox{2.5cm}[0cm][0cm]{(c)}}
    \caption{\textbf{Deterministic photonic QC elements} \textbf{(a)} Qubit logical states obtained with a dual-rail encoding of a pair of photonic channels, \DG{in which single-photon Fock states can be injected in each waveguide and measured at the output through single-photon detectors}. \textbf{(b)} The hopping process of a single photon between two channels can be viewed as a RX gate (see text). \textbf{(c)} Single block of a two-qubit information processing unit: {Multiple repetitions of such an elementary block can be concatenated to obtain the desired two-qubit gate.} Hopping regions (HR) (green boxes) depend on two parameters, ($J_m$, $t_m$), while free propagation (FP) (blue regions) depends only on a single parameter ($t_m$), for a given value of the photon-photon nonlinearity, $U$. The FP layers within the  $\mathcal{U}_{inter}$, $\mathcal{U}_{down}$, and $\mathcal{U}_{up}$ regions are assumed to have fixed propagation time set by the corresponding HR terms, which is implicitly represented by a simple line in the sketch. 
    }
    \label{fig:photonic QC}
\end{figure}

\begin{figure*}[t]
    \centering
    \includegraphics[width=0.32\textwidth]{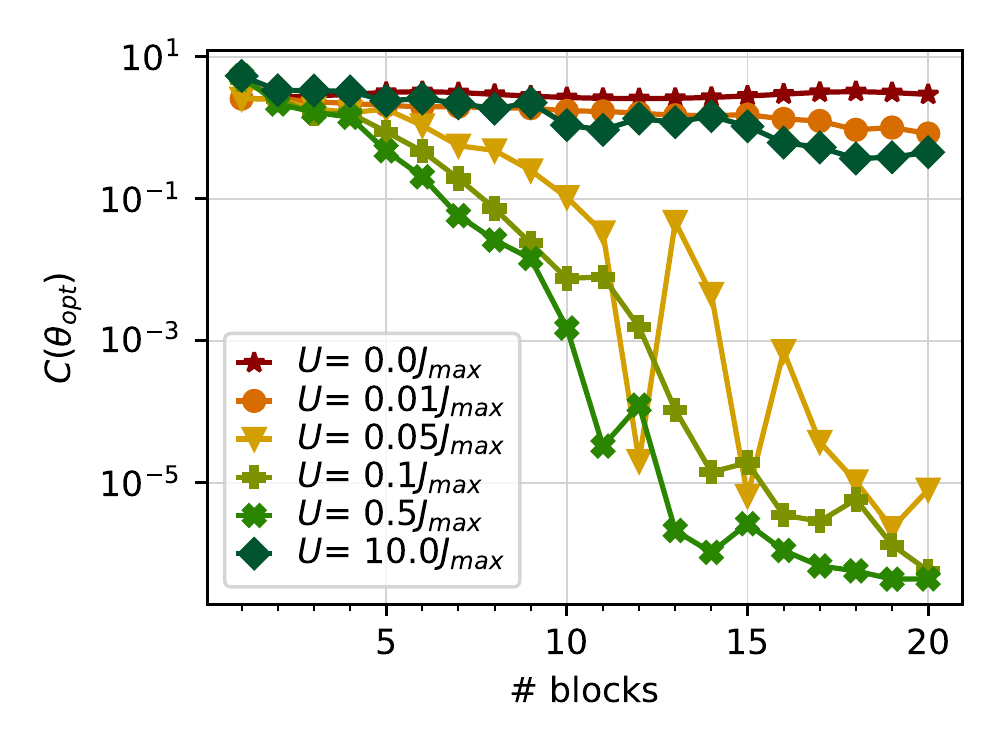}
    \shiftleft{6.cm}{\raisebox{3.6cm}[0cm][0cm]{(a)}}
    \includegraphics[width=0.32\textwidth]{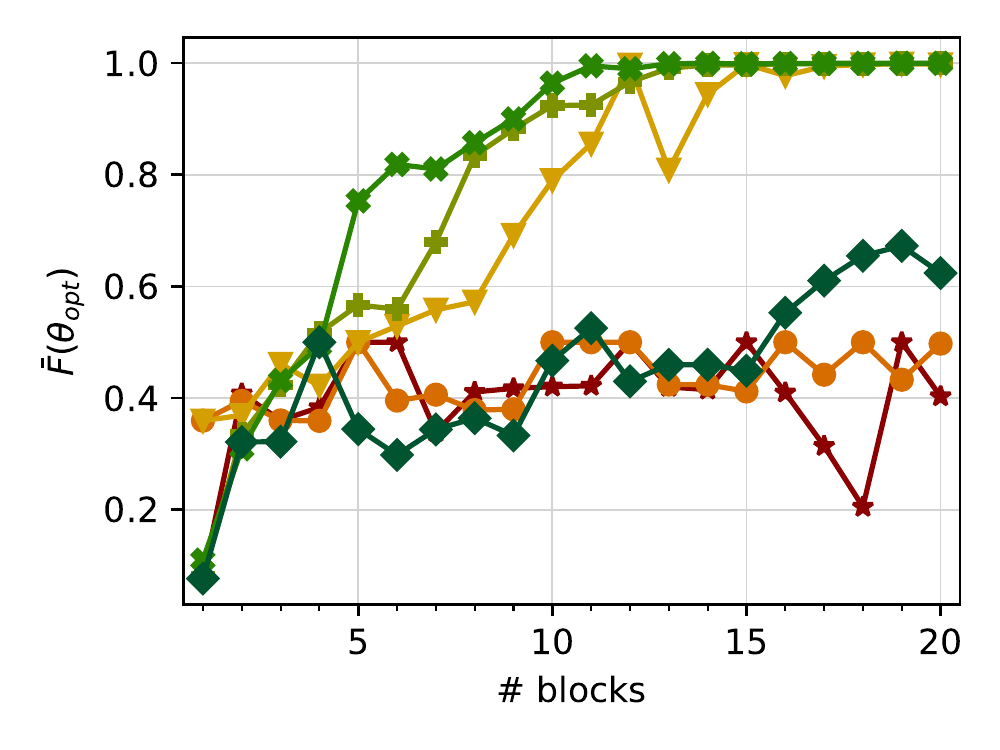}
    \shiftleft{6.cm}{\raisebox{3.6cm}[0cm][0cm]{(b)}}
    \includegraphics[width=0.32\textwidth]{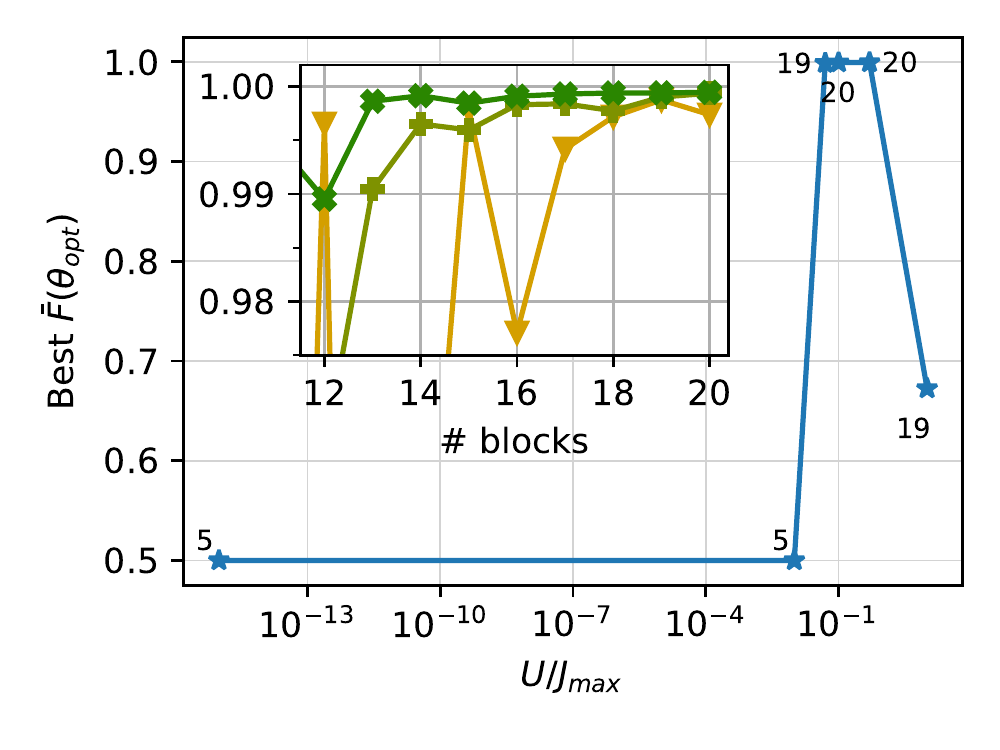}
    \shiftleft{6.cm}{\raisebox{3.6cm}[0cm][0cm]{(c)}}
    \caption{\textbf{CNOT optimization} Comparison of the best cost function (see Eq. \eqref{eq:cost func}) \textbf{(a)} and average gate fidelity (see Eq. \eqref{eq:av gate fidelity}) \textbf{(b)} values for different values of nonlinearity after the optimization of the block structure. Free propagation and interaction times were both fixed to dimensionless values $t=1$. \textbf{(c)} The best average gate fidelity is reported as a function of $U/J_{\mathrm{max}}$. For each point the number of blocks employed to reach the corresponding best value is also reported; the inset shows a close-up of the convergence towards maximal fidelity for $0.05 J_{\mathrm{max}}\leq U\leq 0.5 J_{\mathrm{max}}$.}
    \label{fig:cnot results}
\end{figure*}

{\bf The model}.
The general case of $N$ one-dimensional channels (with corresponding input/output ports) in which quantized photon states propagate {at fixed wave vector $k$}, interfere through {space-dependent} evanescent coupling, and nonlinearly interact when simultaneously present within the same channel can be described by the following Hamiltonian ($\hslash=1$, see Appendix~\ref{appendix:time prop} and~\ref{appendix:operators} for a detailed derivation of this model)
\begin{equation}
\label{eq:ham_n_wg}
\mathcal{H}=\sum_{i=1}^{N}\bigl(\omega_{i} a_{i}^{\dagger}a_i+U_{i} a_{i}^{\dagger 2} a_{i}^2\bigr)
+\frac{1}{2}\sum_{\substack{i,j=1\\ 
j\neq i}}^{N}J_{ij}(x)\bigl(a_{i}^\dagger a_{j}+a_{j}^\dagger a_{i}\bigr)\\,
\end{equation}
with $\mathcal{H}\equiv\mathcal{H}[\{\omega_i\};\,\{U_i\};\,\{J_{ij}(x)\}]$. In Eq.~\eqref{eq:ham_n_wg}, the parameters $\omega_i\equiv\omega_i(k)$ and $U_i\equiv U_i(k)$ denote the energy-momentum dispersion and the Kerr-type nonlinearity in the $i$-th channel, respectively. The space-dependent parameters $\{J_{ij}(x)\}$ denote the hopping terms between next neighboring channels, where $x$ identifies the propagation direction in each one-dimensional (1D) waveguide. In particular, photons in the $i$-th channel are annihilated (created) by bosonic operators $a_{i}\equiv a_{i,k}$ ($a^{\dagger}_{i}\equiv a^{\dagger}_{i,k}$), respectively.\\
In general, the propagation channels might display different features in terms of energy-momentum dispersion as well as photon-photon nonlinearities. For our purposes, we will assume the channels to be identical hereafter. This allows to consider a unique dispersion relation, $\omega=\omega(k)$, and the same nonlinearity in any propagation channel, $U=U_i$. \DG{Generalizations of this scheme are always possible, of course}.

{\bf $N$-channels circuits and qubits encoding}.
When many wave vector components are involved, the characterization of the action of a generic quantum circuit on a given initial many-photon state, $\psi_I$, requires to evolve both in time and space such initial configuration to obtain the final state, $\psi_F$, which contains the QC result.
{In the present case, where only monochromatic single-photon states are considered, the description gets simplified. 
As discussed in Appendix~\ref{appendix:time prop} and~\ref{appendix:operators}, and following Ref.~\cite{Nigro_2022}, one obtains that the action of the generic circuit is encoded into a global unitary operator $\mathcal{U}_{tot}$ such that
\begin{equation}\label{eq:time_evol}
\psi_F = \mathcal{U}_M\, \mathcal{U}_{M-1}\,\cdots \mathcal{U}_{1}\psi_I\equiv \mathcal{U}_{tot}\psi_I \, ,
\end{equation}
with $\mathcal{U}_m\equiv\exp(-i\mathcal{H}[\omega;\,U;\{J^{(m)}_{ij}\}]t_m)$ denoting the unitary propagator in the $m$-th spatial region of the circuit, i.e., for $x\in [x_m,\,x_{m+1}]$, where the inter-channel hoppings can be treated as piecewise constant functions, i.e., $J_{ij}(x)=J^{(m)}_{ij}$. In particular, the parameter $t_m$, with $t_m=(x_{m+1}-x_m)/v_g$, corresponds to the average time spent by photons traveling at group-velocity $v_g$ within the $m$-th sub-region of the circuit.}\\ 
\DG{As a final comment, let us stress that we hereby assume dual-rail encoding to implement single photonic qubits, at difference with Ref.~\cite{Nigro_2022}.} By doing so, a $2N$-channel circuit fed with $N$ single-photon states can be exploited to represent a $N$-qubits quantum state. As sketched in Fig.~\ref{fig:photonic QC}a, the two logical qubit states in a 2-channel device, hereby denoted as $\ket{0}_2$ and $\ket{1}_2$, are defined by the two single-photon Fock states of two adjacent channels, $\ket{1,0}$ and $\ket{0,1}$ respectively, where the former describes a single propagating photon in the upper channel, while the latter describes a single photon propagating in the lower channel. Many-qubits states are straightforwardly obtained by taking the tensor products of the states above in a $2N$-channel device.


{{\bf Single-qubit gates}. Elementary rotations of single qubits can be tailored by suitably combining, e.g., rotations around the $x$- and $z$-axes. As it is shown in the Appendix~\ref{appendix:operators}, for $N=2$ and in the single photon subspace, by means of the dual rail encoding one can directly map the dynamics of two coupled channels into the action of the $RX(\theta)$ rotation gate as}
\begin{equation}
     RX(\theta)=e^{-i\frac{\theta}{2}\sigma_X}=\cos\left(\frac{\theta}{2}\right)\mathds{1}-i\sin\left(\frac{\theta}{2}\right)\sigma_X \quad ,
\end{equation}
multiplied by a global phase factor $e^{-i\omega t}$, where $\sigma_X$ represents the Pauli X-matrix. In particular, as reported in Fig.~\ref{fig:photonic QC}b, the rotation angle, $\theta=2Jt$, depends on the system geometry as well as on the group velocity of the propagating photons. As it will be addressed in the Discussion section, it is reasonable to assume that the simple component depicted in Fig.~\ref{fig:photonic QC}b implements a $RX(\theta)$ gate, with tunable rotation angle.\\
The universal gate set for single-qubit operations can be straightforwardly implemented by adding the $RZ$ gate, obtained, e.g., from the addition of a simple phase shifter in one of the two coupled channels~\cite{Thompson_2011,Wang2020_review}.


{{\bf Two-qubits gates}. It is nowadays widely accepted that deterministic two-qubit gates cannot be devised by means of only linear optical components~\cite{kok_linear_2007,Couteau_2023}. Hence, it is reasonable to argue whether or not the usage of nonlinear components in photonic networks, such as those considered in the present work, could eventually help to achieve this goal. Here we report a theoretical analysis that allows to positively answer this question. In fact, we are going to show that the suitable combination of weakly nonlinear components with linear ones actually represents a key ingredient for the development of a universal QC paradigm based on photonic platforms. Nevertheless, to the best of our knowledge, there is no intuitive way to determine \textit{a priori} which is the optimal arrangement of these elements to deterministically perform a given quantum gate. Therefore, we tackle such a problem by exploiting an inverse design strategy based on a multi-parameter optimization algorithm. In particular, we applied this approach to a class of 4-channel circuits constituted by the concatenation of a finite number of blocks, where each block is defined by combining multiple 2-channels hopping regions with free propagation regions of fixed length, as sketched in Fig.~\ref{fig:photonic QC}c  (see Methods for further details on the definition of the single unitaries).}\\
{Even though several other representations and optimization strategies can be devised, our numerical results suggest that the approach can successfully provide an approximate but highly faithful representation of at least two different entangling gates, namely the controlled-NOT (CNOT) and the M\o{}lmer-S\o{}rensen (M-S). Their matrix representation on the computational basis $S=\{\ket{00}_2, \ket{01}_2, \ket{10}_2, \ket{11}_2 \}$ is explicitly reported in the Methods section for completeness, in Eqs.~\eqref{eq:cnot matrix} and~\eqref{eq:m-s matrix}, respectively. In particular, we show that in both cases it is possible to improve the precision of the representation, which is explicitly quantified by means of the average gate fidelity \fs{$\bar{F}$}, Eq.~\eqref{eq:av gate fidelity}, by simply increasing the number of elementary blocks included in the variational ansatz.
}\\
The optimization procedure is implemented as follows: we first  parametrize a set of possible hopping parameters between the different channels, given in a matrix representation of the unitary evolution operator, i.e., the generalized time propagator in Eq.~\eqref{eq:time_evol}. Then we analyse the performances when increasing the number of sequentially concatenated blocks of the type represented in Fig.~\ref{fig:photonic QC}c, by letting the algorithmic optimizer to minimize the cost function. Details on the choice of the cost function are reported in Methods.

\begin{figure*}[t]
    \centering
    \includegraphics[trim={0 4cm 0 0cm},clip,width=0.9\textwidth]{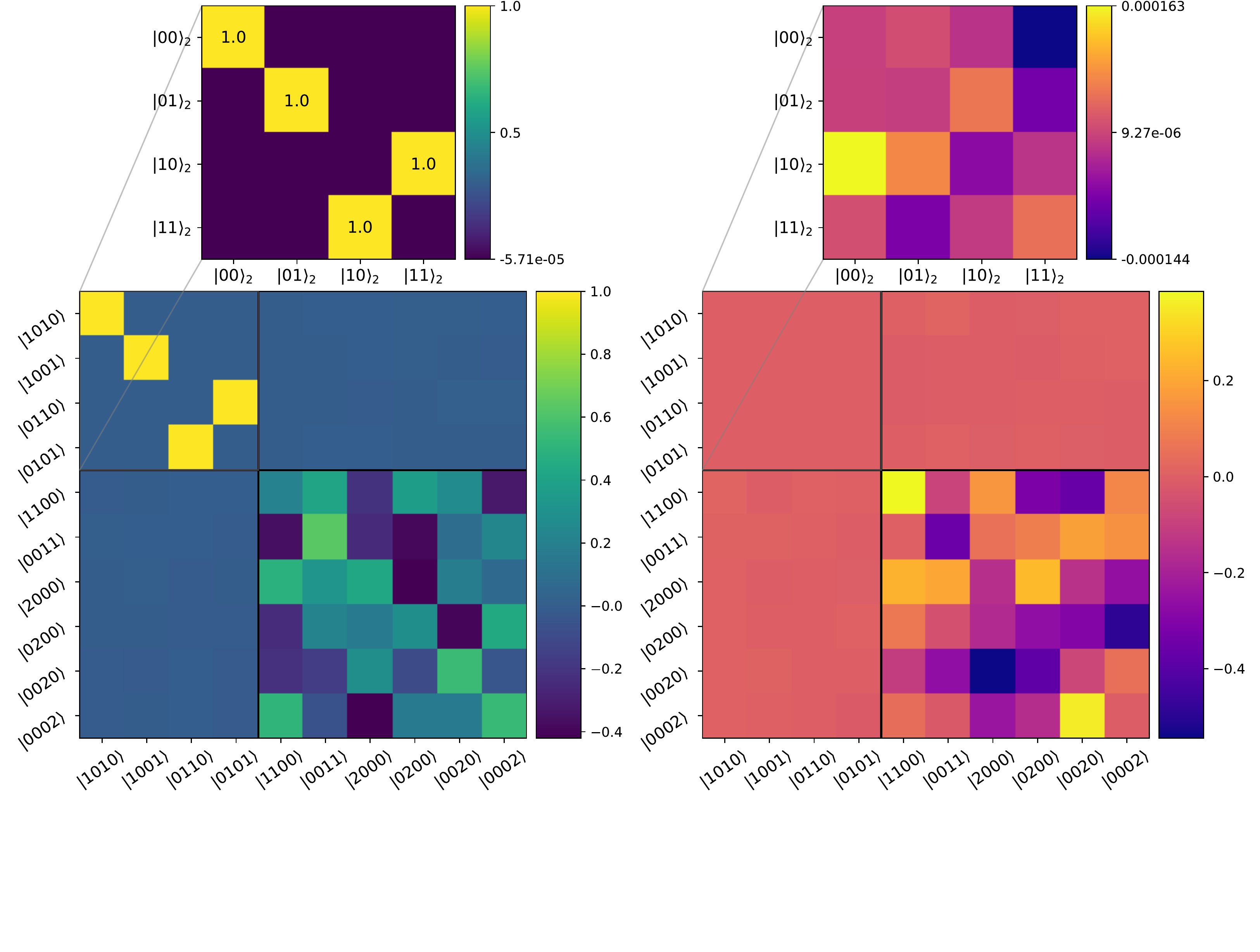}
    \caption{\textbf{Best CNOT matrix} Real and imaginary parts of the approximate CNOT matrix with all the possible two-photon input/output states, when using a 20 blocks structure with $U=0.5J_{\mathrm{max}}$, for which we get $\bar{F}(\theta_{opt})\approx 99.95\%$. Vertical and horizontal black lines divide the logic space from the one based on states that are not in the computational basis. The projection of logic states off the computational space is negligible, as clearly seen from the plot. The zoom highlights the gate matrix in the two-qubit logic space.}
    \label{fig:cnot matrix}
\end{figure*}

{\bf Optimization of the CNOT gate}.
In order to quantify the performances of the optimization scheme, we consider the values of the cost function and the average gate fidelity reached by the multi-parameter optimizer once at convergence, see Eqs.~\eqref{eq:cost func}-\eqref{eq:av gate fidelity} in Methods. Numerical results for such quantities in the case of the CNOT gate are shown in Fig.~\ref{fig:cnot results}a and Fig.~\ref{fig:cnot results}b respectively .
There, we report results for an increasing number of elementary blocks, and for different values of the photon nonlinearity, ranging from zero or very weak to ultra-strong. 
In order to keep the analysis on a general level, we assume the tunneling rate in each hopping region (the HR units in Fig.~\ref{fig:photonic QC}c) as an independent optimization variable such that $0\le J\le J_{\mathrm{max}}$, with $J_{\mathrm{max}}=1$ in dimensionless units, and the propagation time in each sector to be a dimensionless parameter, also fixed as $t=1$. A discussion of the actual dimensions and physical implementations will be given in the following Section.
\DG{As it can be noticed at first glance from Fig.~\ref{fig:cnot results}a, the loss function cannot be minimized for negligible values of the nonlinearity, always displaying values that are significantly larger than zero 
independently of the number of blocks. This behavior is compatible with the well established knowledge that two-qubit entangling gates cannot be realized only by means of linear (or approximately linear) components \cite{kok_linear_2007,Couteau_2023}. On the other hand, we find that this conclusion holds true also in the $U \gg J_{max}$ case, which has to be attributed to the onset of the photon blockade regime between neighboring channels preventing double occupancy of a single channel to a large extent~\cite{Nigro_2022}}. 
However, relatively weak nonlinearities (in particular, ranging between $U/J_{max}=0.05$ and 0.5, in normalized units) give rise to a reasonably fast convergence of the optimizer towards minimal cost function values. {These numerical results suggest that the entanglement between qubit states in such platforms emerges as the result of the competition between hopping (i.e., $J$s) and the entity of the nonlinear shifts affecting two-photon states propagating within the same channel (i.e., $U$). Indeed, even if the latter are never fed into the device as input states, and they do not belong to the computational basis, their excitation does occur during the time evolution within the circuit, and ultimately affects the computation outcome. Thus, the final output of the 4-channel system is the nontrivial result of the extra phase-shifts accumulated in the presence of such nonlinearities.}

\begin{figure*}[ht]
    \centering
    \includegraphics[width=0.49\textwidth]{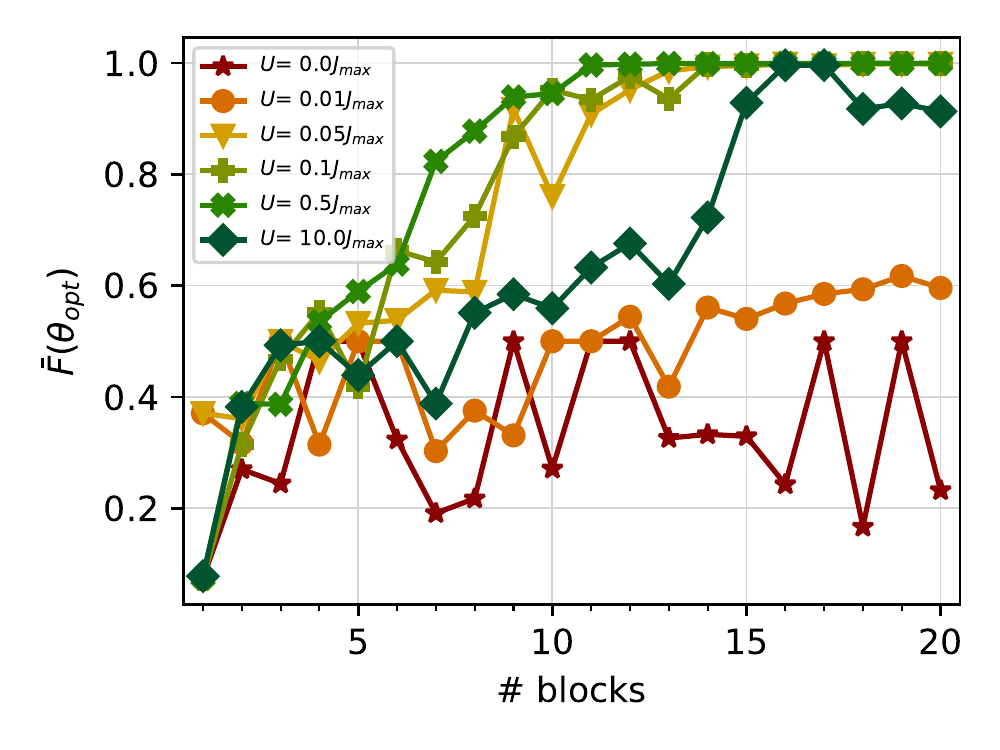}
    \shiftleft{9.cm}{\raisebox{5.5cm}[0cm][0cm]{(a)}}
    \resizebox{0.49\textwidth}{!}{%
    \begin{tikzpicture}
    \node[scale=1.1](a){\includegraphics[width=0.49\textwidth]{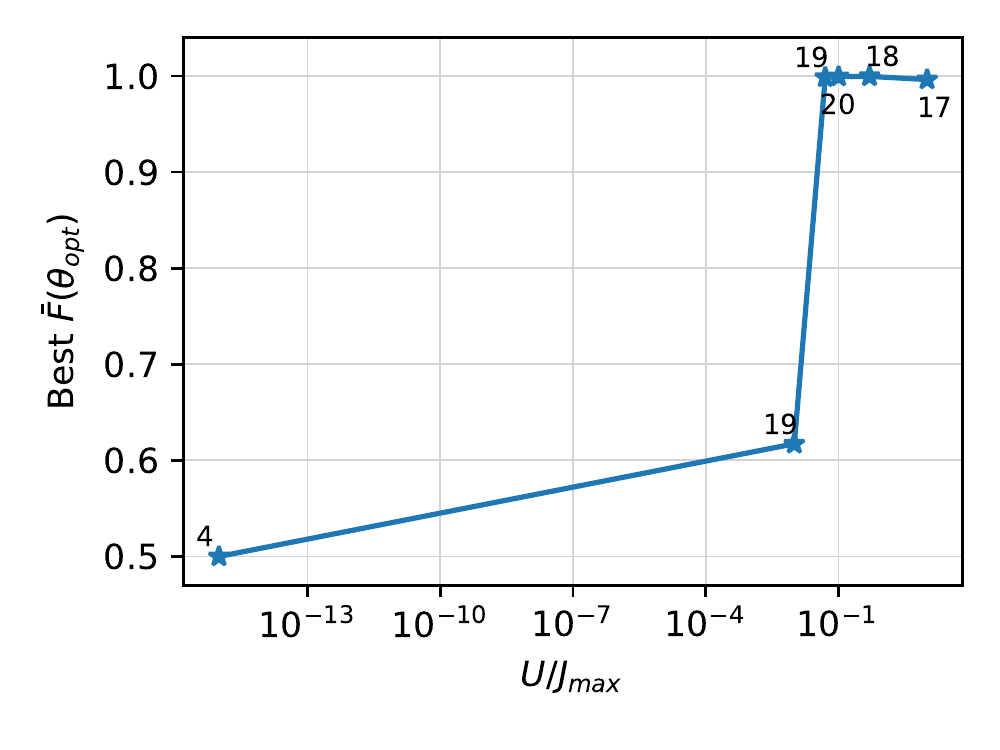}};
    \node at (a.north west)
    [scale=1.1,
    anchor=center,
    xshift=4.48cm,
    yshift=-1.58cm
    ]
    {
        {%
        \setlength{\fboxsep}{0pt}%
          \fbox{\includegraphics[trim={0.1cm .5cm 0.1cm 1.cm},clip,width=0.3\textwidth,]{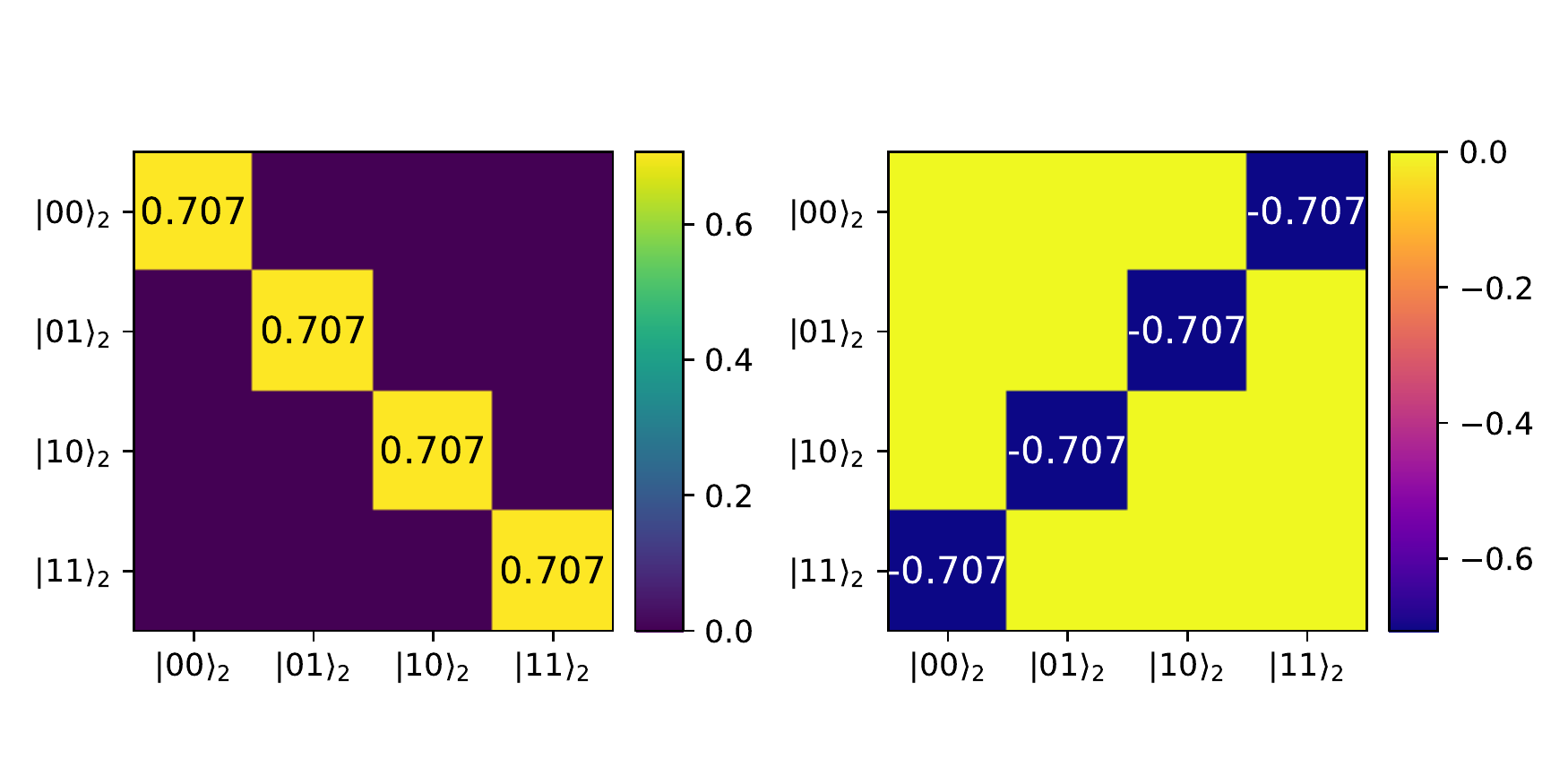}}
          }
    };
    \end{tikzpicture}}
    \shiftleft{8.9cm}{\raisebox{5.6cm}[0cm][0cm]{(b)}}
    \caption{\textbf{M\o{}lmer-S\o{}rensen gate optimization (a)} Best average gate fidelity values comparison after the optimization of the block structure. Free propagation and interaction times were fixed to 1 ps. \textbf{(b)} Plot showing the best average gate fidelity value as a function of $U/J_{\mathrm{max}}$.  In the inset, the optimal M-S matrix is displayed for $U/J_{\mathrm{max}}=0.5$ and 18 blocks, for which we calculate $\bar{F}(\theta_{opt})\approx 99.95\%$.}
    \label{fig:m-s plots}
\end{figure*}

In fact, this is further evidenced by plotting the corresponding average gate fidelity as a function of the number of blocks, in Fig.~\ref{fig:cnot results}b. Interestingly, for nonlinearities in the range $0.05 <U/J_{\mathrm{max}}<0.5$ the gate fidelity reaches values close to 100\% for a number of elementary blocks in the order of 10, although for the lowest value $U/J_{\mathrm{max}}=0.05$ convergence gets slower and oscillating (see also close-up in the inset of Fig.~\ref{fig:cnot results}c). 
To better visualize this behavior, in Fig.~\ref{fig:cnot results}c we report the best gate fidelity results as a function of the nonlinearity, where it seems that the optimal nonlinearity value is $U=0.5 J_{\mathrm{max}}$, which allows obtaining a CNOT gate with high fidelity with reasonably compact structures, i.e., 13 elementary blocks are already sufficient to reach an average gate fidelity very close to 100\% (see also inset).
As an illustration of the actual performances to be expected, the real and imaginary parts of the optimized quantum gate matrix are explicitly shown for the best possible CNOT obtained with our numerical optimization procedure, in Fig.~\ref{fig:cnot matrix} (see upper panels for a close up). The real part seems to perfectly match the CNOT matrix, while the imaginary part is almost irrelevant, thus faithfully reproducing Eq.~\eqref{eq:cnot matrix}. In Fig.~\ref{fig:cnot matrix} we also show the full transfer matrix when taking into account all the possible configurations within the two-photon subspace, which spans, of course, also outside of the logic space of interest (e.g., when two photons simultaneously propagate within the same waveguide channel at the input or output port, respectively). This strongly supports the conclusion that such a gate is deterministic, since no logic state is basically mapped outside the logic space, i.e., with negligibly small amplitude.
Interestingly, it should be noted that \emph{inside} the entangling gate the quantum state can (and must) populate states lying outside the logic space, in order to have the two photons nonlinearly interacting with each other; on the other hand, the output must always be restricted to the logic space for the operation to be defined as \textit{deterministic}. 
The optimal parameters found for this structure are fully reported in a dedicated table in the Appendix~\ref{appendix:opt params}, for a straightforward reproduction of our results from the interested reader.

{\bf Optimization of the M-S gate}.
{While the optimization of the CNOT gate by concatenating a limited number of elementary operations in a weakly nonlinear quantum photonic interferometric circuit is a relevant result per se, here we also show that a similarly efficient algorithmic optimization is possible for other entangling operations on the two-qubits space. Namely, we focus here on the ideal quantum gate defined in Eq.~\eqref{eq:m-s matrix}, i.e. the  M\o{}lmer-S\o{}rensen gate.
The related numerical results are summarized in Fig.~\ref{fig:m-s plots}, where it is shown the behavior of the average gate fidelity as a function of the number of blocks for different values of the nonlinearity (Fig.~\ref{fig:m-s plots}a), as well as the best average gate fidelity obtained for the same values of $U/J_{max}$  (Fig.~\ref{fig:m-s plots}b). Similarly to what was observed for the CNOT gate, it is confirmed as a sort of general trend that convergence towards the ideal operation is reached quite fast for values of the nonlinear parameter in the range  $0.05 J_{\mathrm{max}}<U<0.5 J_{\mathrm{max}}$. This is not the case for negligible or too large nonlinearity values, although the photon blockade for $U/J_{\mathrm{max}}=10$ seems to have less of a detrimental effect here. A summary of this trend is reported in Fig.~\ref{fig:m-s plots}b. The inset of this panel shows the optimal M-S gate obtained for $U/J_{\mathrm{max}}=0.5$, which deterministically implements an almost ideal RXX operation when restricting to the computational basis subspace. A plot of such an optimized M-S gate is explicitly reported in the Appendix~\ref{appendix:extra}. Also in this case, a table reporting the optimal parameters for this structure is given in the Appendix~\ref{appendix:opt params}, for completeness.

\section{Discussion}

\DG{The relevance of the results reported in this work is mainly theoretical. Essentially, we have shown that relatively simple quantum photonic interferometers with weak photon-photon nonlinearities can be optimized to display universal quantum gate operations. However, it is equally relevant to discuss possible physical implementations, and related to that, any possible source of noise, loss, decoherence, fabrication tolerance, which we hereby address at the level of discussion while reporting all the relevant quantitative results in Appendix~\ref{appendix:incoherent dyn} and~\ref{appendix:tolerance}.}

{\bf Realization and material platforms}.
\DG{Several quantum photonic platforms have been put forward in the past few years, mostly based on conventional nonlinear materials in passive semiconductors, such as Si, SiN, glass, etc.~\cite{Wang2020_review}. While the scheme we have presented in this work is general, and can be applied also outside the photonic realm, in principle, here we focus  on discussing its realization in nonlinear photonic circuits. First, we notice that single-photon Fock states can be produced at high repetition rates and high purity levels from single quantum emitters \cite{Somaschi2016}, as well as detected at the output of the device through highly efficient single-photon detectors \cite{Chang2021-single_photon_dect}. In addition, injection of single-photon Fock states generated from quantum emitters into photonic integrated circuits does not constitute a technological bottleneck, nowadays \cite{Loredo_2017_BSampling}.} 

\DG{Then, we notice that in view of devising a full QC platform the optimized circuits implementing two-qubit operations must be complemented with isolated two-channel circuits in which single-qubit rotations are performed. In this respect, the hopping rate $J$ can be tuned by suitably choosing the spatial separation between two adjacent channels. On the other hand, the parameter $v_g$ can be tuned and externally controlled by properly shifting the working point along the photon dispersion. This task can be practically achieved, for instance, by local temperature tuning, which results in local refractive index changes. 
This would ensure, for instance, the flexibility to perform $RX$ rotations with externally controlled rotation angles.
In principle, the application of the optimization procedure described in this work might lead to the targeted design of a quantum photonic integrated circuit fulfilling the requirements to achieve deterministic quantum gate operations for each specific platform under consideration.} 

\DG{In terms of possible material platforms, besides the well-established silicon-on-insulator technology, where remarkable advances in quantum photonic experiments have been recently shown~\cite{Clementi2023}, a viable example might be the SiN platform, for which the high nonlinearity, low-loss, and mature fabrication of complex circuits~\cite{Xiang2022_review} might turn out to be an optimal combination to realize a proof of principle demonstration, at least. }
\DG{More recently, exciton-polaritons in semiconductor nanostructures have been proposed as an interesting platform where quantum photonic applications could be realized~\cite{Ghosh2020,Nigro_2022}, relying on the first experimental evidence for the quantum nature of the propagating polariton field excited from a single-photon Fock state~\cite{Cuevas_2018}, a key requirement of our theoretical scheme. The polariton nonlinearity is of the Kerr-type, as it is well established \cite{Verger2006,Delteil2019,Munoz-Matutano2019}, and it allows to realize the model Hamiltonian in Eq.~\eqref{eq:ham_n_wg} with values of $U$ that are compatible with the ones required for the optimal operations identified in the previous Section. In fact, by assuming, e.g., typical excitation energies in the $\hbar\omega~1$ eV range, hopping parameters can be assumed such that $\hbar J_{\mathrm{max}}=1$ meV, as an order of magnitude estimate \cite{Beierlein2021}. 
Then, the single-photon nonlinearity depends on the actual field confinement \cite{Verger2006,Ferretti2012}, and for 1D propagating polariton wave-packets with spatial extension of the order of their wavelength it may range from 10 $\mu$eV (realistic, see, e.g., Refs.~\cite{Delteil2019}) up to 100 $\mu$eV depending on the material (optimistic, including recent claims for enhanced nonlinearities of dipolar polaritons \cite{Rosenberg2018,Togan2018,Suarez-Forero2021}). These values would then correspond to $U/J_{\mathrm{max}}=0.01 - 0.1$, i.e., in the range where we have shown that two-qubit gates can actually be realized with large fidelity,  in Figs.~\ref{fig:cnot results} and \ref{fig:m-s plots}. In addition, $J_{\mathrm{max}}$ can also be reduced in realistic samples (depending on the distance between neighbouring waveguide channels), thus making the optimal parameter ($U/J_{\mathrm{max}}=0.5$) within reach. This would also allow to reduce the number of blocks required to achieve an optimal fidelity close to 100\%, which is particularly relevant when losses have to be included in the discussion, as it will be addressed in the next paragraph. } 

{\bf Losses, decoherence, fabrication imperfections}.
\DG{We have introduced a scheme based on the implementation of a unitary Hamiltonian evolution, but realization of this paradigm in state-of-art photonic devices has to inevitably cope with the effects of losses and decoherence. Moreover, tolerance of the simulated figures of merit, such as gate fidelities, to fabrication imperfections or fluctuations of structure parameters has to be assessed. Here we provide a short discussion summarizing the main conclusions about these effects, as obtained from extensive numerical simulations within an open quantum system approach. The details are reported in the Appendix~\ref{appendix:incoherent dyn} and~\ref{appendix:tolerance}.\\
First, we address the issue of population losses. As already analysed before \cite{Nigro_2022}, propagation losses may affect the actual signal intensity detected at the output of the device, but correlations between single-photons properly normalized to the detected intensity are not affected by such loss mechanism. Also in this specific case, whenever the output signal is detected, the transfer operation is automatically realized with high fidelity, thus making the gate operations fully deterministic anyways. 
In fact, provided the whole interferometer length remains within the propagation lifetime, the quantum gate operation should be preserved (with a given efficiency due to the loss of signal). To check this conjecture numerically, we have solved the quantum master equation for our model, including population losses in a Lindblad term (see Appendix~\ref{appendix:incoherent dyn}). The outcome is strikingly straightforward: the absolute gate fidelity has an exactly exponential decay with a lifetime given by the inverse of the two-photon population decay rate, that is $\tau=1/(2\gamma)$. Thus, by normalizing the fidelity to such exponential factor, the close-to-ideal fidelity is recovered. The effects of population decay can thus be fully mitigated by proper ``renormalization'' of the output signal for a given decay rate $\gamma$.\\
Other sources of noise, such as thermal noise or pure dephasing, can also be quantified through dissipative terms in the master equation (see Appendix~\ref{appendix:incoherent dyn}). First, photon number fluctuations can be considered absolutely negligible, given that low working temperatures in the Kelvin range produce negligible thermal photons in the visible/near-infrared range, and assuming pure Fock-state injection guarantees no presence of higher photon number states (as it would be the case, e.g., in an attenuated laser source). On the other hand, the effects of number-dependent dephasing might have an impact on the relative coherence between the qubit basis states. In order to properly quantify these effects, we have considered a number-dependent Lindblad term with a pure dephasing rate $\gamma_{deph}$. Our numerical results (reported in  Appendix~\ref{appendix:incoherent dyn}) suggest that a pure dephasing rate $\gamma_{\mathrm{deph}}<10^{-2} \gamma$ has no practical effects on the qubits decoherence, thus preserving the overall fidelity of the two-qubit entangling operation. On the other hand, dual rail encoding naturally allows to minimize the effects of the loss of relative coherence between the two logical states of a single qubit, which are defined from single photons propagating in each channel. The results are interesting per se, and they might serve as a guide for experimentalists working on specific material platforms, once a proper characterization of dephasing rates is performed. \\
Finally, we have also considered the effects of static fluctuations of the model parameters on the overall gate performances, as it might occur when fabrication imperfections or structural disorder come into play after the circuit realization. Also in this case, by considering realistic parameters fluctuations in state of the art semiconductor technology it is concluded that the optimal gate fidelities predicted in this work can be safely preserved against static noise.\\
To conclude this Section we try to be specifically more quantitative on the estimation of loss parameters for prospective polariton integrated circuits, and see where we actually stand in terms of achievable results. First, let us notice that intrinsic   exciton polaritons lifetime in inorganic semiconductor nanostructures has been measured in the order of 100 ps \cite{Nelsen2013}. Moreover, exploiting the concept of bound-state in the continuum in a patterned waveguide geometry to suppress out-of-plane radiation losses, lifetimes in the order of 300 ps have been measured in polariton condensates (see, e.g., additional material of Ref.~\cite{Ardizzone2022}). Recently, such propagation lifetimes have been also confirmed in a polariton waveguide geometry \cite{Suarez-Forero2021}. Based on these results, it is very likely that combining suitably engineered photonic lattices with single-quantum well samples, propagating single-polariton states with large group velocities and ultra-long propagation lengths might be engineered \cite{Zanotti2022}. With these numbers in mind, we can estimate the single polariton decay rate in the $\hbar\gamma\sim 1-10$ $\mu$eV range, i.e., same order of magnitude expected for $\hbar U$. For the results shown, e.g., in Figs.~\ref{fig:cnot results} and \ref{fig:m-s plots}, a sequence of 15 blocks with an average duration of about 8 ps per block (i.e., assuming a 1 ps average propagation time in each of the 8 sectors schematically represented in Fig.~\ref{fig:photonic QC}c) amounts to an estimated total propagation time of 120 ps when considering the realistic $U/J_{\mathrm{max}}=0.05$ case. This would allow to achieve a two-qubit entangling gate with fidelity in excess of 99.5\% in a propagation time well within the polariton lifetime of, e.g., 200 ps.
}

{\bf Summary}.
We have proposed a quantum computing model based on the realization of a universal gate set of qubit operations in nonlinear photonic interferometers, where a dual-rail type of qubit encoding is assumed. We have shown that weak single-photon nonlinearities within the same propagating channel allow to build robust deterministic entangling gates between two such photonic qubits with high average gate fidelity, whose quest has been one of the major issues in the field for years. The optimal realization of this operation on-chip is achieved by a suitable concatenation of something between 10 and 20 elementary blocks, each containing all the possible combinations of propagation unitaries defined on a 4-port device{, without the need for additional ancillary waveguides}.
{On the quantitative side, we have shown that optimal CNOT and M-S quantum gates can be designed with  99.95\% theoretical fidelities.
It is worth noting, for comparison, that currently available QC devices have state-of-the art CNOT fidelities in the order of 99.77\% with superconducting circuit architectures~\cite{Kandala_PRL2021}, and M-S fidelities of 99.6\% with trapped ion few qubits devices\footnote{Latest data from the IonQ Aria QPU specifications, see, e.g., https://ionq.com/posts/july-25-2022-ionq-aria-part-one-practical-performance}. }
Finally, our optimal operations have been tested against the main sources of loss, thermal noise, pure dephasing, also showing a good resilience to static parameters fluctuations derived from, e.g., fabrication imperfections in actual devices.
In conclusion, we believe these results might foster further research towards the realization of quantum devices to be used as building blocks of a canonical model of quantum computation employing single propagating photons as information carriers.

\section{Methods}

{\bf Two qubit gates}.
In this work we have targeted two-qubit operations defined as controlled-NOT (CNOT) and M\o{}lmer-S\o{}rensen (M-S)~\cite{molmer-sorensen1,molmer-sorensen}, which are paradigmatic entangling quantum gates. In particular, the CNOT is described by the following ideal operation in matrix representation on the two-qubits basis~\cite{nielsen00}

\begin{align}
    \label{eq:cnot matrix}
    CNOT =& \begin{bmatrix}
        1 & 0 & 0 & 0\\
        0 & 1 & 0 & 0\\
        0 & 0 & 0 & 1\\
        0 & 0 & 1 & 0\\
    \end{bmatrix} \quad .
\end{align}
Its action is such that the state of the second (target) qubit is flipped when the first (control) qubit is in its logical state $\ket{1}_2$. \\
The M-S gate~\cite{molmer-sorensen} is an alternative entangling operation consisting of a RXX gate with a fixed angle of $\pi/2$. 
Differently from the CNOT, the M-S gate has a non-trivial imaginary part:

\begin{align}
\label{eq:m-s matrix}
    RXX(\pi/2) = \frac{1}{\sqrt{2}}&\begin{bmatrix}
        1 & 0 & 0 & -i\\
        0 & 1 & -i & 0\\
        0 & -i & 1 & 0\\
        -i & 0 & 0 & 1\\
    \end{bmatrix} \quad ,
\end{align}
where $RXX(2\theta)=\exp\left(-i\theta\sigma_X\otimes\sigma_X\right)$.\\

In the present implementation these two gates are realized by exploring the time propagation of a pair of single-photon states into a 4-port quantum photonic interferometer.
The general Hamiltonian describing such a system is Eq.~\eqref{eq:ham_n_wg} with $N=4$, which accounts for two main phenomena, i.e., two-photon nonlinear phase shifts and hopping of photons between adjacent channels, respectively. Hence, all possible time-evolution unitary operators can be written as a tensor product of a single channel operator, $\mathcal{U}_{FP}$, accounting for nonlinear propagation, and a two-channel one describing hopping events, $\mathcal{U}_{HR}$.
The matrix form of these two operators is reported in the Appendix~\ref{appendix:operators}, where the explicit dependence on the parameters of the Hamiltonian model, as well as on the propagation time in each sector, can be appreciated.\\
In light of these considerations, these two matrices can be used to define the set of elementary 4-channel operations, that is \{$\mathcal{U}_{free}$, $\mathcal{U}_{paral}$, $\mathcal{U}_{inter}$, $\mathcal{U}_{down}$, $\mathcal{U}_{up}$\}, needed for the parametrization of the fundamental block depicted in Fig.~\ref{fig:photonic QC}c. In particular, their explicit expressions read

\begin{align}
    \label{eq:F}
    & \mathcal{U}_{free} =\mathcal{U}_{FP}\otimes\mathcal{U}_{FP}\otimes\mathcal{U}_{FP}\otimes\mathcal{U}_{FP}\\
    \label{eq:I_paral}
    & \mathcal{U}_{paral} = \mathcal{U}_{HR} \otimes \mathcal{U}_{HR} \\
    \label{eq:I_inter}
    & \mathcal{U}_{inter} = \mathcal{U}_{FP}\otimes \mathcal{U}_{HR}\otimes \mathcal{U}_{FP}\\
     \label{eq:I_down}
    & \mathcal{U}_{down} = \mathcal{U}_{FP}\otimes \mathcal{U}_{FP}\otimes \mathcal{U}_{HR}\\
    \label{eq:I_uo}
    & \mathcal{U}_{up} = \mathcal{U}_{HR}\otimes \mathcal{U}_{FP}\otimes \mathcal{U}_{FP}
\end{align}
where $\otimes$ denotes the tensor product. For our purposes, it is worth noticing that the two $\mathcal{U}_{HR}$ operators in Eq.~\eqref{eq:I_paral} are defined, in general, with different values of $J_{ij}$. This degree of freedom is exploited in the optimization procedure. Once the unitary operator $\mathcal{U}_b$ describing the single block depicted in Fig~\ref{fig:photonic QC}c is parametrized, the total time-propagator $\mathcal{U}_{tot}$ for a structure with $M$ blocks is obtained by considering the ordered product of such block operators, that is 
\begin{equation}
  \mathcal{U}_{tot}(\{\theta_s\})=\bigotimes_{b=1}^M \mathcal{U}_b=\mathcal{U}_M\,\mathcal{U}_{M-1}\cdots\mathcal{U}_2\,\mathcal{U}_1,  
\end{equation}
 where $\{\theta_s\}$ denotes the set of physical parameters used for representing the $M$-block system.

{\bf Cost function, minimization scheme, and gate fidelity}. In this section we briefly describe the main ingredients used in the optimization scheme, namely the cost function, the numerical optimizer, and the average gate fidelity used to assess the gate performances described in the previous sections.\\
The cost function considered in the present work is defined as 
\begin{equation}
    \label{eq:cost func}
    C(\{\theta_s\}) = \sum_{i,j\in S} |\mathcal{U}_{tot}(\{\theta_s\})_{i,j}-\mathcal{T}_{i,j}|^2 \, 
\end{equation}
where $S=\{\ket{00}_2, \ket{01}_2, \ket{10}_2, \ket{11}_2 \}$ is the computational basis, while $\mathcal{U}_{tot}(\{\theta_s\})$ and $\mathcal{T}$ denote the total unitary operator describing the 4-channel system and the target ideal operation (either the CNOT or the M-S in Eqs.~\eqref{eq:cnot matrix}-\eqref{eq:m-s matrix}), respectively.\\
By definition, the cost function is a non-negative quantity, and it should go to zero only if the parametrization provided by $\mathcal{U}_{tot}$ is exact. As a consequence, the optimization procedure aims at finding the best approximation of the target operator $\mathcal{T}$ by looking for the set of values of the physical parameters $\{\theta_s\}$ that minimizes $C(\{\theta_s\})$. In practice, this task is performed by means of numerical routines.
In particular, we make use of the Scipy~\cite{scipy} implementation of L-BFGS-B~\cite{byrd1995limited, zhu1997algorithm} optimizer (whose execution has been accelerated with JAX~\cite{jax2018github, jaxopt_implicit_diff}), which is a limited-memory algorithm for solving large nonlinear optimization problems subject to simple bounds on the variables.\\
Once at convergence, the routine returns the optimal set of physical parameters $\{\theta^{opt}_{s}\}$ that miminimizes $C(\{\theta_s\})$ for a given value of the nonlinearity and number of blocks.
The actual performances of the optimization procedure are subsequently quantified by computing the \textit{average gate fidelity}, $\bar{F}(\{\theta^{opt}_{s}\})$. The explicit expression of this figure of merit reads 
\begin{equation}
    \label{eq:av gate fidelity}
    \bar{F}(\{\theta^{opt}_{s}\}) = \frac{1}{|S|} \sum_{i\in S} |\langle i|\mathcal{U}_{tot}^\dagger(\{\theta^{opt}_{s}\}) \mathcal{T}|i\rangle|^2 \, .
\end{equation}
Similarly to what reported above, the matrix $\mathcal{T}$ in Eq.~\eqref{eq:av gate fidelity} denotes one of the two target operators, as defined, e.g., in Eqs.~\eqref{eq:cnot matrix}-\eqref{eq:m-s matrix}. Consequently, the particular values  $\{\theta^{opt}_{s}\}$ depend explicitly on the chosen $\mathcal{T}$. 
An explicit derivation of Eq.~\eqref{eq:av gate fidelity} is reported in Appendix~\ref{appendix:gate fidelity}. \\
For each numerical result reported in the manuscript (corresponding to a given value of  nonlinearity $U$ and to a given number of blocks), we have sampled 20 different initial sets of hopping parameters $\{J_{ij}\}$ recorded as one-dimensional vectors, and 20 different optimization procedure are then executed. Among them we select the best final configuration in terms of the achieved accuracy, i.e. the one leading to the minimal value of the cost function. The initial set of hopping parameters is sampled from a Gaussian distribution centered in $0.5J_{\mathrm{max}}$ with standard deviation $0.1J_{\mathrm{max}}$.\\
As a final comment, we notice that it might sound appealing to try using the average gate fidelity as a cost function. However, since this quantity is only sensitive to the squared modulus of the overlap amplitudes, it cannot be actually used to optimize $\mathcal{U}_{tot}(\{\theta_s\})$. Indeed, if on the one hand, it is easy to show that   
\begin{equation}
    C(\{\theta^{opt}_{s}\})=0 \Rightarrow \bar{F}(\{\theta^{opt}_{s}\}) = 1 \, ,
\end{equation}
on the other hand, the converse statement does not hold true, in general. In fact, real and imaginary parts may compensate each other to maximize $ |\langle i|\mathcal{U}_{tot}^\dagger(\{\theta^{opt}_{s}\}) \mathcal{T}|i\rangle|^2$. In other words, there exist sets of parameter values that maximize the fidelity, without simultaneously minimizing  $C(\{\theta_{s}\})$.

\section*{Data Availability}
All the data and simulations that support the findings of this study are available from the corresponding author upon reasonable request.

\section*{Code Availability}
The Python codes developed for this study are available from the corresponding author upon reasonable request.

\section*{Contributions}
D.G. conceived the original idea, F.S. developed the code and performed the simulations, D.N. and D.G. supervised the work. All authors contributed to manuscript writing and discussions about the results.

\section*{Competing Interests}
The authors declare no competing interests. 

\section*{Acknowledgements}
This research was supported by the Italian Ministry of Research (MUR) through PRIN 2017 project INPhoPOL. D.G. acknowledges the PNRR MUR project CN00000013 - National Research Center on ``HPC, Big Data and Quantum Computing'' (HPC), F.S. and D.N. acknowledge the PNRR MUR project PE0000023 - National Quantum Science Technology Institute (NQSTI). The authors acknowledge useful scientific discussions with D. Bajoni, F. Giorgino, E. Maggiolini. DG acknowledges V. Savona for inspiring discussions.




\bibliography{main}

\appendix

\onecolumngrid

\section{Time-propagators of multi-channel systems}
\label{appendix:time prop}
In this section we provide further details concerning the formalism adopted in the Results section of the main text and we show how to explicitly derive Eq.~(2).\\
For what concerns this first issue, let us consider as an example the case where the system is made of only two propagation channels. Results for the $N$-channel case are a straightforward generalization of what is discussed below.\\
In this case, the general model reported into the main text reduces to the following
\begin{equation}
    \mathcal{H}=\bigl(\omega a_{1}^{\dagger}a_1+U a_{1}^{\dagger 2} a_{1}^2\bigr)+\bigl(\omega a_{2}^{\dagger}a_2+U a_{2}^{\dagger 2} a_{2}^2\bigr)+J(x)\bigl(a_{1}^\dagger a_{2}+a_{2}^\dagger a_{1}\bigr),
\end{equation}
where the operator $a^{(\dagger)}_{j}$ annihilates (creates) a photon in the $j$-th channel, $\omega$ is the energy-momentum of photons and $U$ denotes the photon-photon nonlinearity. Contrary to the general case, in the present one the coupling between the two channels can be parametrized by means of a single space-dependent function $J(x)$ only determined by the physical distance between the two channels at position $x$ along the propagation direction. In addition, since such coupling is of evanescent nature, the function $J(x)$ is expected to decay exponentially when increasing the separation between the two channels.\\
Let us suppose now to have $M=N+Z$ different regions in our system: $N$ regions where the channels run in parallel and are sufficiently close so that here $J(x)\neq 0$, and $Z$ spatial subregions where $J(x)\simeq 0$. If this is the case, it is therefore reasonable that the space-dependent coupling $J(x)$ can be approximated by means of $M$ piecewise-constant functions, that is by means of the following 
\begin{equation}
J(x)\simeq\sum_{m=1}^M J^{(m)} \chi_m(x) \quad \, \chi_m(x)=\left\{
\begin{array}{lc}
1  & x\in [x_{m-1},x_m]\\
0  & \mbox{otherwise}
\end{array}\right.      
\end{equation}
with $\{J^{(m)}\}$ denoting the set of constants values assumed by $J(x)$ in the $M$ spatial subregions $[x_{m-1},x_m]$ mentioned above. By assuming that single photons propagate across the structure at the group-velocity $v_g$ specified by $\omega=\omega(k)$, the action of a generic 2-channel circuit on a generic multi-photon state $\psi_I$ is given by
\begin{equation}\label{eq:Utot_def}
\psi_F = \mathcal{U}_M(t_M)\, \mathcal{U}_{M-1}(t_{M-1})\,\cdots \mathcal{U}_{1}(t_{1})\psi_I\equiv \mathcal{U}_{tot}\psi_I,
\end{equation}
where $\mathcal{U}_m(t_m)\equiv\exp(-i\mathcal{H}[\omega;\,U;\{J^{(m)}\}]t_m)$ denotes the unitary propagator in the subregion $x_{m-1}<x<x_m$, and with $t_m=(x_m-x_{m-1})/v_g$. Eq.~\eqref{eq:Utot_def} can be considered as a definition of $\mathcal{U}_{tot}$, that is the unitary operator describing the action of the 2-channel structure on a generic state.\\ 

\section{The $\mathcal{U}_{FP}$ operator, the $RX(\theta)$ gate, and the $\mathcal{U}_{HR}$ operator}
\label{appendix:operators}
In this section we discuss how to derive the explicit expression of the $\mathcal{U}_{FP}$, $RX(\theta)$ and $\mathcal{U}_{HR}$ operators.\\
As a starting point of our discussion, we observe that $\mathcal{U}_{FP}$ is a single-channel operator, while the latter two correspond to two-channel transformations. On top of that, the $RX(\theta)$ operator is a single-qubit transformation, while the $\mathcal{U}_{HR}$ accounts, in general, for the case where one or more particles are propagating in adjacent channels, so it describes the evolution of quantum configurations that cannot always be mapped into single-qubit states. This means that such three operators are represented by matrices having different sizes. Notice, however, that we are interested in describing scenarios where at most two single-photon states are injected within the whole 4-channel system (two-qubit case). This means that (i) the relevant single-channel states for our purposes are given by $\{\vert 0\rangle,\,\vert 1\rangle,\,\vert 2\rangle\}$, and that (ii) the description of two photons propagating into a $N$-channel platform requires at most $3^N$ configurations, corresponding to the set of states $\{\vert 0\rangle,\,\vert 1\rangle,\,\vert 2\rangle\}^{\otimes N}$, where 
\begin{equation}
     \{\vert 0\rangle,\,\vert 1\rangle,\,\vert 2\rangle\}^{\otimes N}=\bigotimes^N_{i=1}\{\ket{0},\ket{1},\ket{2}\}_i
\end{equation}
and with
\begin{equation}
    \{\vert 0\rangle,\,\vert 1\rangle,\,\vert 2\rangle\}^{\otimes 2}=\{\vert 0\rangle,\,\vert 1\rangle,\,\vert 2\rangle\}\otimes \{\vert 0\rangle,\,\vert 1\rangle,\,\vert 2\rangle\}=\{\ket{0,0},\ket{0,1},\ket{0,2},\ket{1,0},\ket{1,1},\ket{1,2},\ket{2,0},\ket{2,1},\ket{2,2}\}
\end{equation}
denoting the tensor product of $ \{\vert 0\rangle,\,\vert 1\rangle,\,\vert 2\rangle\}$ with itself (which gives the basis set we used to describe a two-channel system).

With all these considerations in mind, given that 
\begin{equation}\label{eq:creation_annihilation_action}
a^{\dagger}\vert n\rangle =\sqrt{n+1} \vert n+1\rangle,\quad a\vert n\rangle=\sqrt{n}\vert n-1\rangle,
\end{equation}
it is straightforward to obtain for instance $\mathcal{U}_{FP}$. Indeed, since for $N=1$ the Hamiltonian $\mathcal{H}$ is diagonal, one has
\begin{equation}
\mathcal{U}_{FP} =e^{-i\mathcal{H}t}= 
\begin{bmatrix}
    1 & 0               & 0\\
    0 & e^{-i\omega t} & 0\\
    0 & 0               & e^{-i2(U+\omega)t}\\
\end{bmatrix}
\end{equation}
where we assumed particles to propagate on a region of length $l=v_g t$.\\

As in the previous section, let us now consider the general model in Eq.~(1) of the main text restricted to the case $N=2$. Depending on the number of particles that are assumed to propagate within the structure, one ends up with the expression of $RX(\theta)$ or $\mathcal{U}_{HR}$. When considering a single particle propagating in the structure, the Hamiltonian simplifies to 
\begin{equation}
    \mathcal{H}=\omega a_{1}^{\dagger}a_1+\omega a_{2}^{\dagger}a_2+J(x)\bigl(a_{1}^\dagger a_{2}+a_{2}^\dagger a_{1}\bigr).
\end{equation}
In particular, in every region where the hopping coupling is constant, that is $J(x)=J$, by expanding such operator onto the single-photon basis $\{\vert 1,0\rangle,\,\vert 0,1\rangle\}$, one has that its exponential reads
\begin{equation}
\label{appendix:1pol-dyn}
    \mathcal{U}^{(1)}_{HR}(t)=e^{-i\mathcal{H}t}=e^{-i\omega t}(\cos(Jt)\mathds{1}-i\sin(Jt)\sigma_x)\equiv  RX(2Jt),
\end{equation}
with $\mathds{1}$ and $\sigma_x$ being the identity and the Pauli-X matrices respectively. The superscript $``(1)"$ in Eq. \eqref{appendix:1pol-dyn} is used to stress that such expression holds true only in the single-photon subspace.\\

For what concerns the expression of $\mathcal{U}_{HR}$, since the Hamiltonian preserves the number of particles, its structure can be determined by exploring the different subspaces with fixed photon number. In the single-photon sector, one obtains the operator reported in Eq. \eqref{appendix:1pol-dyn}. In a similar fashion, into the two-photon subspace spanned by $\{\vert 1,1\rangle,\vert 0,2\rangle, \vert 2,0\rangle\}$, one obtains after a little algebra that 
\begin{equation}
\label{appendix eq:2pol-dyn}
\mathcal{U}_{HR}^{(2)}   = 
\begin{bmatrix}
A & B & B \\
B & C & D \\
B & D & C
\end{bmatrix}
\end{equation}
with
\begin{align*}
    & A = \biggl(\cos{\Omega t}+i\frac{\sin{\Omega t}}{\Omega}U\biggr)e^{-i\phi t}\ ,\\
    & B = -i\frac{\sin{\Omega t}}{\Omega}\sqrt{
    2}J e^{-i\phi t}\ ,\\
    & C = \biggl\{ \frac{1}{2\Omega}[-iU\sin{\Omega t}+\Omega\cos{\Omega t}]+\frac{1}{2}[\cos{Ut}-i\sin{U t}]\biggr\}e^{-i\phi t}\ ,\\
    & D = \biggl\{ \frac{1}{2\Omega}[-iU\sin{\Omega t}+\Omega\cos{\Omega t}] -\frac{1}{2}[\cos{Ut}-i\sin{U t}]\biggr\}e^{-i\phi t}\ ,\\
\end{align*} 
where $\Omega = \sqrt{4J^2+U^2}$ and $\phi=U+2\omega$.\\

The final expression of $\mathcal{U}_{HR}$ is obtained by combining the results reported in Eq. \eqref{appendix:1pol-dyn} and \eqref{appendix eq:2pol-dyn}, with those describing the dynamics in the 3- and 4-particles sectors. However, even if the coefficients related to these latter Fock subspaces appear in the final form of $\mathcal{U}_{HR}$, we do stress that they are not involved in the system dynamics considered in the main text, that is when at most 2-photons are injected in the system. In particular, by using the following ordering prescription
\begin{equation}
    \vert a,\,b\rangle \rightarrow i(a,\,b)=a\cdot 3+(b+1) \ ,
\end{equation}
which associates an integer $i(a,\,b)\in [1,\,9]$ to each two-channel state $\vert a,\,b\rangle$ of the basis set $\{\vert 0\rangle,\,\vert 1\rangle,\,\vert 2\rangle\}^{\otimes 2}$ (so that $\vert 0,0\rangle$ corresponds to the first row/column, $\vert 0,1\rangle$ to the second row etc), one obtains that $\mathcal{U}_{HR}$ is represented by the following $9 \times 9$ operator:
\begin{equation}
\label{eq:U_IR global}
\mathcal{U}_{HR}   = 
\begin{tikzpicture}[baseline=-\the\dimexpr\fontdimen22\textfont2\relax ]
\matrix (m)[matrix of math nodes,left delimiter={[},right delimiter={]}]
{
    1 & 0 & 0 & 0 & 0 & 0 & 0 & 0 & 0 \\ 
    0 & E & 0 & F & 0 & 0 & 0 & 0 & 0\\
    0 & 0 & C & 0 & B & 0 & D & 0 & 0\\
    0 & F & 0 & E & 0 & 0 & 0 & 0 & 0\\
    0 & 0 & B & 0 & A & 0 & B & 0 & 0\\
    0 & 0 & 0 & 0 & 0 & G & 0 & H & 0 \\ 
    0 & 0 & D & 0 & B & 0 & C & 0 & 0\\
    0 & 0 & 0 & 0 & 0 & H & 0 & G & 0 \\ 
    0 & 0 & 0 & 0 & 0 & 0 & 0 & 0 & I \\ 
};
\begin{pgfonlayer}{myback}
\highlight[blue]{m-2-2}{m-2-2}
\highlight[blue]{m-4-4}{m-4-4}
\highlight[blue]{m-2-4}{m-2-4}
\highlight[blue]{m-4-2}{m-4-2}

\highlight[orange]{m-3-3}{m-3-3}
\highlight[orange]{m-7-7}{m-7-7}
\highlight[orange]{m-3-7}{m-3-7}
\highlight[orange]{m-7-3}{m-7-3}
\highlight[orange]{m-5-5}{m-5-5}
\highlight[orange]{m-3-5}{m-3-5}
\highlight[orange]{m-5-3}{m-5-3}
\highlight[orange]{m-5-7}{m-5-7}
\highlight[orange]{m-7-5}{m-7-5}
\end{pgfonlayer}
\end{tikzpicture}
\end{equation}

where
\begin{align*}
    &E=\cos{(Jt)}e^{-i\omega t} \ ,\\   
    &F=-i\sin{(Jt)}e^{-i\omega t}\,\\
    &G=\cos{(2Jt)}e^{-i(3\omega+2U)t }\,\\
    &H=-i\sin{(2Jt)}e^{-i(3\omega+2U)t}\,\\
    &I=e^{-4(\omega + U)t}
\end{align*}
where the highlighted blocks correspond to coefficients coming from single- (blue) and two-photon (orange) dynamics, respectively.

\section{Derivation of average gate fidelity expression}
\label{appendix:gate fidelity}
{Given two density operators $\rho$ and $\sigma$ representing valid quantum states such that
\begin{equation}\label{eq:density_matrix_props}
    X=X^{\dagger},\,X\geq 0 ,\, \Tr{X}=1,\, \quad X=\rho,\,\sigma \, ,
\end{equation}
a measure of the distance between such states is provided by the \textit{fidelity}, whose expression reads \cite{Jozsa_1994}:
\begin{equation}
    \label{appendix eq:fidelity}
    F(\rho, \sigma) = \biggl(\Tr\bigl[ \sqrt{\rho^{1/2}\sigma\rho^{1/2}}\bigr] \biggr)^2 \, ,
\end{equation}
with $F(\rho,\,\sigma)=F(\sigma,\,\rho)$. In general, the evaluation of such quantity requires the computation of the square root of either $\rho$, or $\sigma$. This task is simplified whenever pure states are involved, i.e., density operators expressed as 
\begin{equation}
    \rho=\ket {\psi_{\rho}} \bra{\psi_{\rho}} \quad \mbox{and/or}\quad \sigma=\ket {\psi_{\sigma}} \bra{\psi_{\sigma}} \, .
\end{equation}
For the sake of clarity, let us suppose $\rho=\ket {\psi_{\rho}} \bra{\psi_{\rho}}$. Since in this case $\sqrt{\rho}=\rho$ (due to the purity of $\rho$), the expression of $F(\rho,\,\sigma)$ reads
\begin{equation}
    F(\rho,\,\sigma)=\biggl(\Tr\bigl[ \sqrt{\rho^{1/2}\sigma\rho^{1/2}}\bigr] \biggr)^2=\biggl(\Tr\bigl[ \sqrt{\rho\sigma\rho}\bigr] \biggr)^2=\biggl(\Tr\bigl[ \sqrt{\ket {\psi_{\rho}} \bra{\psi_{\rho}} \sigma\ket {\psi_{\rho}} \bra{\psi_{\rho}} }\bigr] \biggr)^2=\bra{\psi_{\rho}} \sigma \ket{\psi_{\rho}} \, ,
\end{equation}
where the last equality follows from the linearity of the trace operation. Similarly, when $\sigma$ represents a pure state, for symmetry reasons the fidelity can also be written as 
\begin{equation}\label{eq:fidelity_rho_vs_pure_sigma}
    F(\rho,\sigma)=\bra{\psi_{\sigma}} \rho \ket{\psi_{\sigma}} \, .
\end{equation}
In particular, this is the case whenever the quantum states involved in the evaluation of the fidelity are obtained by means of two evolution operators defined, e.g., $A$ and $B$, each acting on an initial pure state such as  
\begin{equation}
    \rho_A= A\ket {\psi^{(0)}_{\rho}} \bra{\psi^{(0)}_{\rho}}A^\dagger \quad , \quad \sigma_B=B\ket {\psi^{(0)}_{\sigma}} \bra{\psi^{(0)}_{\sigma}} B^\dagger \, ,
\end{equation}
for which the fidelity is rewritten as
\begin{equation}\label{eq:two_pure_states}
    F(\rho_A,\,\sigma_B)=\vert\bra{\psi^{(0)}_{\rho}} A^{\dagger}B \ket{\psi^{(0)}_{\sigma}}\vert^2  {=|\langle\psi^{(t)}_{\rho}|\psi^{(t)}_{\sigma}\rangle|^2} \, .
\end{equation}
Then, the \textit{average gate fidelity} expression, whose results are shown in the main text, follows from Eq.~\eqref{eq:two_pure_states} by considering $\ket{\psi^{(0)}_{\rho}}=\ket{\psi^{(0)}_{\sigma}}$, and it corresponds to an average over the contributions associated to each state in the two-qubit computational basis, with $A$ and $B$ denoting the transfer operators associated to the optimized quantum circuit $\mathcal{U}$ and the target operator (i.e, CNOT or RXX gates in our case), respectively.
}


\section{Coherent vs. incoherent dynamics}
\label{appendix:incoherent dyn}
{The results reported in the main text rely on the assumption that the quantum circuit is composed of ideal lossless components, i.e., implementing unitary transformations between input and output quantum states. Here, we provide further theoretical considerations to address the relevant issue of possible deviations induced by the coupling of the platform to the surrounding environment, which is unavoidably present in real experiments. In particular, we focus on three different types of coupling and their effects on the quantum circuit: particle losses, thermal noise, and pure dephasing. Such a task is accomplished by means of the Lindblad master equation formalism for open quantum systems, see e.g.~\cite{breuer2002theory}. In the unitary regime, the time evolution is prescribed by the Schr\"{o}dinger equation and all the relevant information is encoded into the Hamiltonian $\mathcal{H}$ and the quantum state $\ket{\psi}$, which completely specify the system evolution. An alternative picture is provided by the von Neumann differential equation for the system density operator $\rho$, which for the unitary case (i.e., equivalent to the Schr\"{o}dinger equation) is  formally expressed as 
\begin{equation}
    i \hslash \partial_t \rho =  [\mathcal{H}, \rho] \, .
\end{equation}
In the Lindblad formalism, the presence of a coupling to the environment is taken into account by including extra terms into the von Neumann equation. In fact, the master equation governing the time evolution of the density operator generally reads
\begin{equation}\label{eq:generic_Lindblad}
   i \hslash \partial_t \rho = [\mathcal{H}, \rho]+ i \hslash \sum_{j=1}^{N_{\mathrm{ch}}}\gamma_j D[O_j, \rho] \, ,
\end{equation}
in which
\begin{equation}\label{eq:generic_dissipator}
    D[\hat{O}, \rho] = \hat{O} \rho \hat{O}^\dagger - \frac{1}{2}\left(\hat{O}^\dagger \hat{O}, \rho +\rho \hat{O}^\dagger \hat{O}\right) \, .
\end{equation}
Equation \eqref{eq:generic_Lindblad} describes a general scenario where the system is perturbed by a number $N_{\mathrm{ch}}$ of different noise sources, each one characterized by its own rate and system Lindblad operator, i.e., $\gamma_j$ and $\hat{O}_j$ respectively.\\
\subsection{Particle loss and thermal noise}\label{sec:particle_loss_VS_thermal}
In this theoretical framework, terms leading to particle losses and thermal noise have the same origin and can be regarded as the two sides of the same coin. Usually, the former is pictured by means of Lindblad operators that destroy single-particle excitations in the system. Similarly, thermal noise terms account for the finite probability of having an excitation injected into the system due to the thermal mean-occupation associated with the configuration of the environment, thus controlled by Lindblad operators that create an excitation in the open quantum system. Therefore, the most reasonable choice is to assume the two effects to be controlled by annihilation and creation operators of the system, in the present case given by the set of $\{a_j\}$ and $\{a^{\dagger}_j\}$, where the $j$ index labels the propagation channel. The rates at which single quanta can escape or are injected within the system do depend on both the spectral properties of the quantum circuit, i.e., theh energy-momentum dispersion of the propagation channels, as well as the environment properties. When a system is  primarily coupled to the surroundings through radiative interactions (as it is the case, e.g., in polariton systems), the environment can be treated as a continuum of free-propagating electromagnetic modes at finite temperature, $T_{env}$, thus represented by an ensemble where a mode at energy $E=\hslash \omega$ contains a mean-number of photons given by the Bose distribution, that is
\begin{equation}\label{eq:bose_number}
    n_{B}(\omega,\,T_{\mathrm{env}})=\frac{1}{e^{\hbar\omega /k_B T_{\mathrm{env}}}-1},
\end{equation}
where $k_B$ is the Boltzmann constant. In particular, as extensively discussed in Ref.~\cite{breuer2002theory}, one can express rates associated to particle losses and thermal noise affecting the system occupation at energy $\hbar \omega$ through Eq.~\ref{eq:bose_number}. More explicitly, one obtains that
\begin{equation}\label{eq:loss_rate}
    \gamma_{-}(\omega,\,T_{env})=\gamma (n_{B}(\omega,\,T_{env})+1) \, ,
\end{equation}
and 
\begin{equation}\label{eq:thermal_rate}
    \gamma_{+}(\omega,\,T_{env})=\gamma n_{B}(\omega,\,T_{env}) \, ,
\end{equation}
where $\gamma$ denotes the zero-temperature particle loss rate, and $\gamma_{-}$ and $\gamma_{+}$ describe the rates at which a single-particle excitation at energy $\hslash\omega$ is lost ($-$) or injected ($+$, thermal noise) in the system due to the finite temperature occupation of the environment. In particular, whenever 
\begin{equation}\label{eq:energy_vs_temperature}
    \hslash \omega \gg k_B T_{env},
\end{equation}
$n_{B}(\omega,\,T_{env})\to 0$ and the expression above can be approximated by their zero-temperature limits, i.e.,
\begin{equation}\label{eq:zero_temperature_loss}
    \gamma_{-}(\omega,T_{env}\to 0)\to\gamma,
\end{equation}
and 
\begin{equation}\label{eq:zero_temperature_thermal_noise}
     \gamma_{+}(\omega,T_{env}\to 0)\to 0.
\end{equation}
 We notice that at temperatures in the range of few Kelvin, that is $T_{\mathrm{env}}\simeq 1-10$ K, the typical energy scale introduced by thermal fluctuations is $k_B T_{\mathrm{env}}\simeq 10^{-4}-10^{-3}$ eV. Therefore, whenever the quantum system is excited at an energy scale in the eV range (as assumed in the main text), the zero-temperature limit can be considered as an excellent approximation. In fact, such a condition is usually met in standard exciton-polariton experiments, where the working point on the energy-momentum dispersion is chosen slightly below the quantum-well exciton resonance, usually located at the eV scale in standard materials, and the samples are kept at constant temperature $T_{\mathrm{env}}$ in the few K range (see, e.g., \cite{Cuevas_2018,Suarez-Forero2021,Ardizzone2022}). In other words, as far as the typical scenarios considered in this manuscript, thermal fluctuations can be safely neglected and noise terms introduced by particle loss in the channel $j$ can be sketched by means of the following operator
 \begin{equation}\label{eq:dissipator_particle_loss}
     D_{\mathrm{Loss}}[a_j,\,\rho]=a_j \rho a_j^{\dagger}-\frac{1}{2}\left(a^{\dagger}_j a_j \rho + \rho a^{\dagger}_j a_j \right) \, ,
 \end{equation}
 with a single-particle loss rate given by the zero temperature limit, $\gamma$, in each propagation channel (we assume identical  propagation channels, without loss of generality).\\
 In order to understand whether the presence of single-particle loss might affect the results discussed in the main text, we consider the effects of a dissipator term as in Eq.~\eqref{eq:dissipator_particle_loss} on a system characterized by the Hamiltonian obtained by numerical optimization of the quantum circuit parameters. In particular, since the main target of the present analysis is to show how such terms might affect the quantum circuit response when fed with two single-photon states, i.e., defining the computational basis states in the dual-rail encoding, we first pay attention to the behavior of the average occupation of such states at the end of the quantum circuit as a function of $\gamma$. More explicitly, results reported below have been obtained by means of the following protocol. At $t=0$, we consider a pure density-operator (input state) defined as 
 \begin{equation}\label{eq:input_states_comp_basis}
     \rho_{0}=\ket{\psi^{(0)}}\bra{\psi^{(0)}},\quad \ket{\psi^{(0)}}\in \{\cket{0,\,0},\,\cket{0,1},\,\cket{1,0},\,\cket{1,1}\}.
 \end{equation}
The given state is then evolved in time with the Lindblad master equation reported in Eq.~\eqref{eq:generic_Lindblad} (in particular, by means of an explicit Runge-Kutta 4(5) scheme), and the output state at {$\rho_{\mathrm{OUT}}\equiv\rho(t=t_{\mathrm{tot}})$} 
is then analyzed and information about the results of the computation are extracted. We stress that the particular form of $\rho_{\mathrm{OUT}}$ depends on  $\rho_{(0)}$.\\
 A few results obtained for the quantum circuit implementing the CNOT gate are reported in Fig. \ref{fig:example_CNOT_lossy}, in particular, the one with the highest average gate fidelity obtained with our optimization scheme. In Fig.~\ref{fig:example_CNOT_lossy} we report the behavior of the populations of the output state $\rho_{OUT}$ (i.e., diagonal entries), as a function of the ratio $\gamma/\gamma_0$, in which we define $\gamma_0=1/t_{\mathrm{tot}}$ (where $t_{\mathrm{tot}}$ is the total propagation time of the quantum circuit), for the different input states reported in Eq.~\eqref{eq:input_states_comp_basis}. This choice is motivated by the expression of the fidelity reported in Eq.~\eqref{eq:fidelity_rho_vs_pure_sigma}, which connects the average gate fidelity to output populations. Indeed, when the state $\sigma=\ket{CNOT(\psi^{(0))})}\bra{(\psi^{(0))})CNOT}$ describes the action of the ideal CNOT gate on the input state $\rho_{(0)}$ corresponding to one of the computational basis states, the output state can be defined as
 \begin{equation}
     \ket{\psi^{CNOT}}\bra{\psi^{CNOT}}= \ket{CNOT(\psi^{(0)})}\bra{(\psi^{(0)})CNOT},
 \end{equation}
we can express the fidelity with respect to the numerically calculated output state in the lossy quantum circuit as  
 \begin{equation}\label{eq:fidelity_out_VS_CNOT}
     F_{\gamma}(\rho_{\mathrm{OUT}} \ ,\ \sigma)= \bra{\psi^{CNOT}}\rho_{OUT}\ket{\psi^{CNOT}} \, ,
 \end{equation}
 which is the population in $\rho_{\mathrm{OUT}}$ associated to $\ket{\psi^{CNOT}}$.
 Due to the high average fidelity, in the absence of single-particle losses ($\gamma/\gamma_0=0$), the optimized structure behaves essentially as the ideal CNOT gate, with an almost perfect unit transfer of population from the input state to the output state (IN $\to$ OUT, as detailed in the legend of Fig.~\ref{fig:example_CNOT_lossy}a). When the  $\gamma/\gamma_0$ ratio increases, the output population decreases accordingly. In particular, numerical results show that the very same trend in population decay is observed for all the input density operators representing the four states in the computational basis. In addition, in agreement with the discussion provided in Ref.~\cite{Nigro_2022}, such a decay is compatible with an exponential decay solely depending on the number of particles in $\rho(t=0)$, which is 2 in our case (as evidently shown by the dashed line in Fig.~\ref{fig:example_CNOT_lossy}a),  
and on the ratio $\gamma/\gamma_0$.\\
 \begin{figure}
    \includegraphics[width=0.49\textwidth]{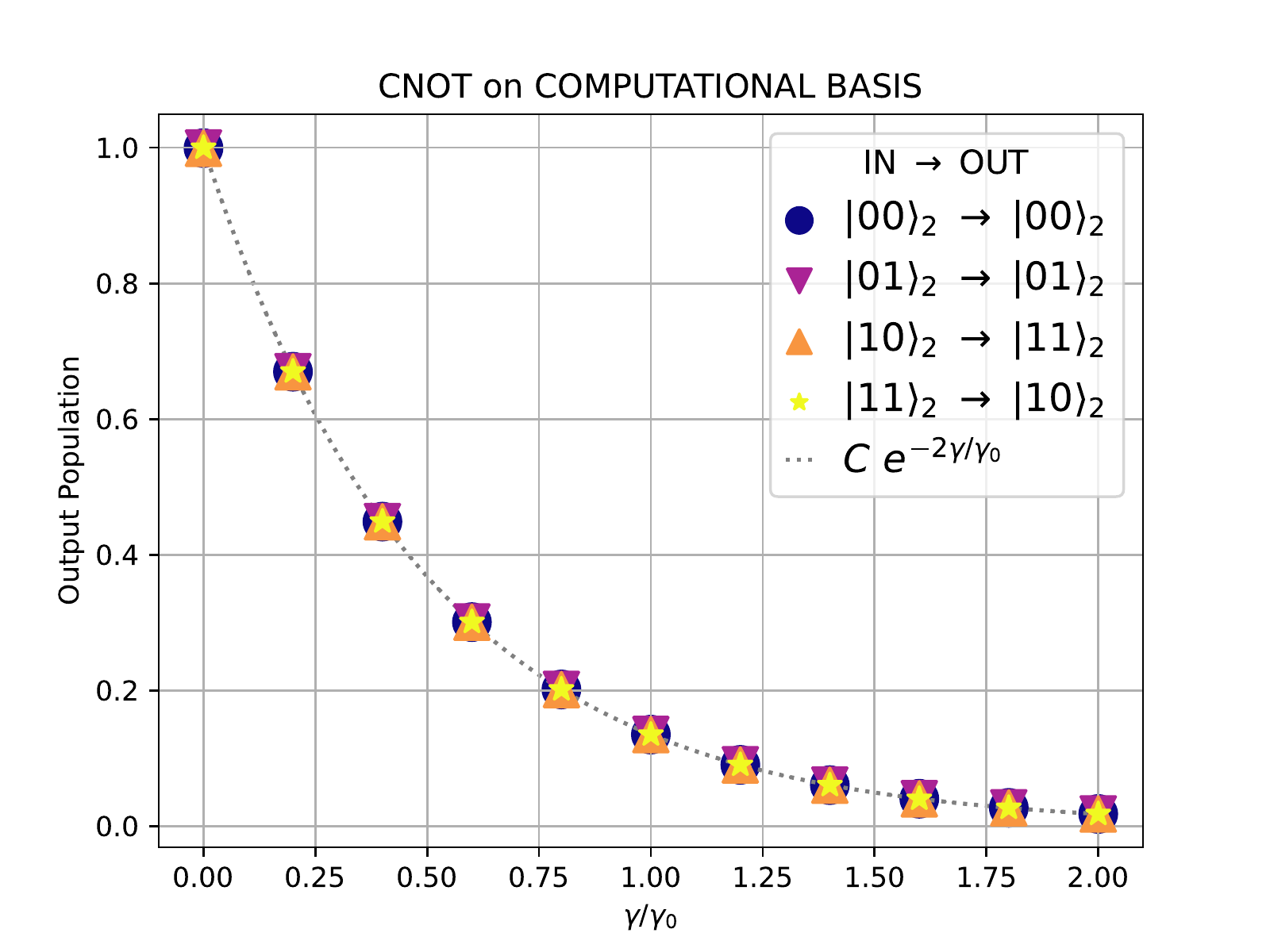}
    \shiftleft{9.cm}{\raisebox{5cm}[0cm][0cm]{(a)}}
    \includegraphics[width=0.49\textwidth]{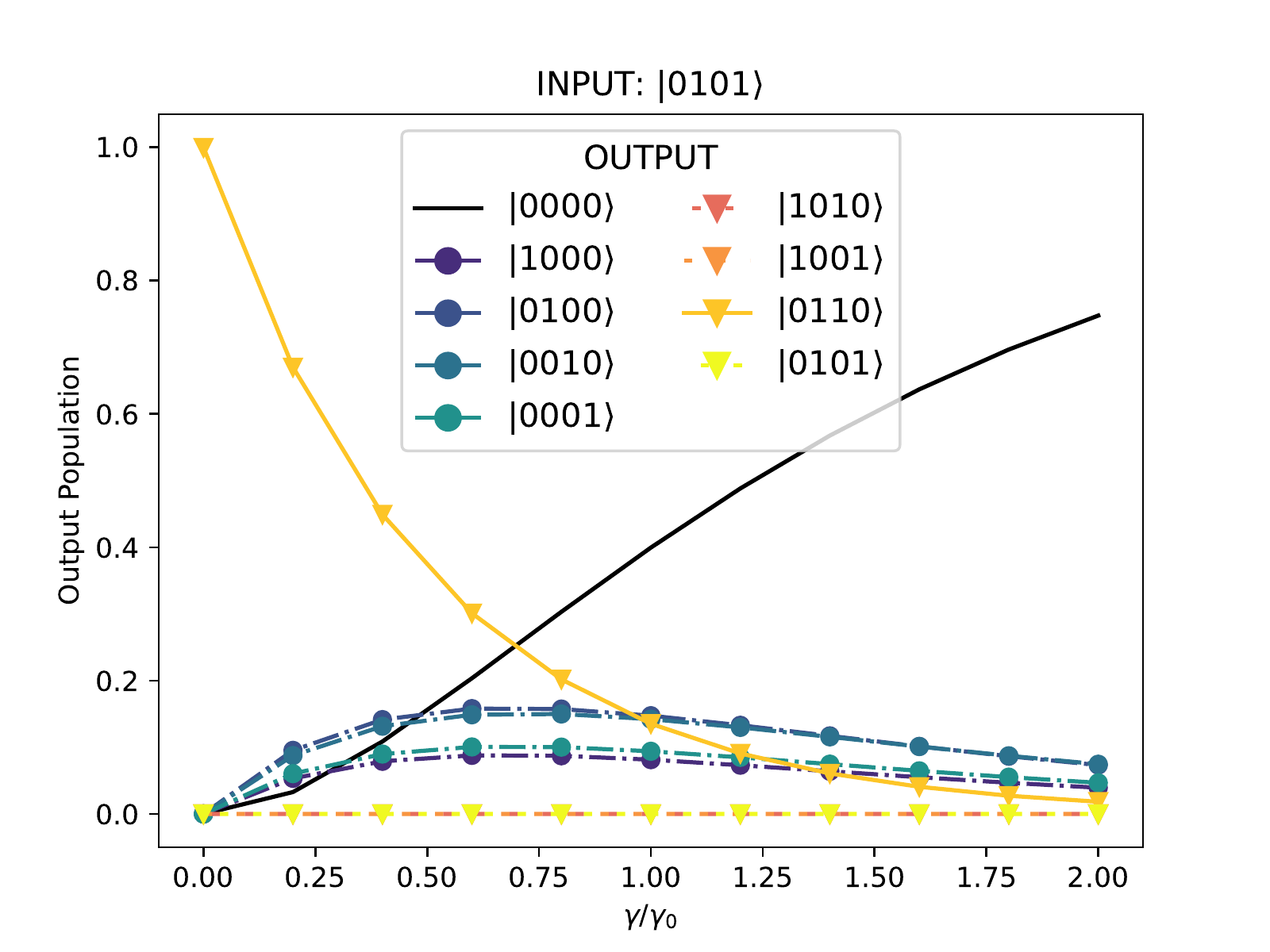}
    \shiftleft{9cm}{\raisebox{5cm}[0cm][0cm]{(b)}}
    \caption{Numerical results showing the gate fidelity for the optimized quantum circuit implementing the CNOT gate in the presence of single-particle loss, assuming here {the best architecture optimized, i.e. for $U/J_{\mathrm{max}}=0.5$ and 20 blocks (without loss of generality)}. We explicitly show (a) the behavior of the numerically calculated output state population (OUT) on increasing $\gamma/\gamma_0$, for any initial state preparation belonging to the computational basis (IN), see the legend, and plot of the exponentially decaying function for the fidelity. The explicit expression of the parameter $C$ is given by $C=\vert \bra{\psi^{(0)}}\mathcal{U}^{\dagger}_{tot}\{\theta_s^{opt}\} \ket{\psi^{CNOT}}\vert^2$. In the present case, $C\simeq 1$ for all the input states. Then, we also show (b) the behavior of the populations in all the basis states (see the legend) on increasing $\gamma/\gamma_0$. Data in this panel correspond to an initial input state given by $\rho(t=0)=\cket{1,1} \cbra{1,1}=\ket{0,1,0,1}\bra{0,1,0,1}$, and the output produced by the ideal CNOT is $\cket{1,0}=\cket{0,1,1,0}$; in the legend, output states in the left column are outside the computational basis, while the ones on the right column are the 4 computational basis states.}
    \label{fig:example_CNOT_lossy}
\end{figure}
Furthermore, as confirmed by the results  reported in Fig.~\ref{fig:example_CNOT_lossy}b, in the presence of single-particle loss the only two-particle component in $\rho_{\mathbf{OUT}}$ is exactly the one expected for the ideal CNOT gate (i.e., for $\gamma=0$). Specifically, in this panel we report data obtained when the initial pure state is given by 
\begin{equation}
    \rho(t=0)=\cket{1,1} \cbra{1,1}=\ket{0,1,0,1}\bra{0,1,0,1}.
\end{equation}
In the unitary regime, such configuration is mapped with the highest fidelity into the computational basis state $\cket{1,0}=\ket{0,1,1,0}$, and any other basis state remains empty. In particular, this happens for single-particle states (namely, $\ket{1,0,0,0}$,$\ket{1,0,0,0}$, $\ket{0,0,1,0}$ and $\ket{0,0,0,1}$), as well as any other state belonging to the computational basis. On increasing the $\gamma$, the population of the initial configuration is not perfectly transferred to $\cket{1,0}$, as it is evident from the plot in Fig.~\ref{fig:example_CNOT_lossy}b, in which other states other than $\cket{1,0}=\ket{0,1,1,0}$ become populated, but only belonging to the vacuum or single-particle sector. As already mentioned, the crucial point here is that the other states belonging to the two-particle sector and defining the computational basis remain unpopulated. Similar behaviors are observed when considering as $\rho_{(0)}$ a pure state corresponding to the other computational basis states (not shown). Hence, this trend in numerical data suggests that the average gate fidelity in the presence of losses can be expressed as  
 \begin{equation}\label{eq:fidelity_with_losses}
     \bar{F}_{\gamma\neq 0}(\{\theta_{opt}\})=\bar{F}_{\gamma=0}(\{\theta_{opt}\})e^{-2\gamma/\gamma_0} \, ,
 \end{equation}
 where $\bar{F}_{\gamma=0}(\{\theta_{opt}\})=\bar{F}(\{\theta_{opt}\})$ corresponds to the average gate fidelity used in the main text for the given $U/J_{\mathrm{max}}$ value. Conversely, we can state that from  Eq.~\eqref{eq:fidelity_with_losses} certainly follows the general expression for the ``renormalized'' output fidelity 
 \begin{equation}
      \bar{F}(\{\theta_{opt}\})=\bar{F}_{\gamma}(\{\theta_{opt}\})e^{2\gamma/\gamma_0} \, .
 \end{equation}
 In summary, if on the one hand particle losses affect the absolute gate efficiency, on the other hand, once the output signal is properly normalized, our results suggest that the quantum circuit still behaves like a CNOT gate on the computational basis, i.e., pairs of photons emerging from the structure correspond always to the output state produced by the ideal CNOT gate, as requested from the deterministic computational protocol: any time two-photons are simultaneously detected via single-photon detectors at the output ports of the 4-channel interferometer, the two qubit gate is realized with almost 100\% fidelity.\\
 \begin{figure}
    \centering
    \includegraphics[width=0.49\textwidth]{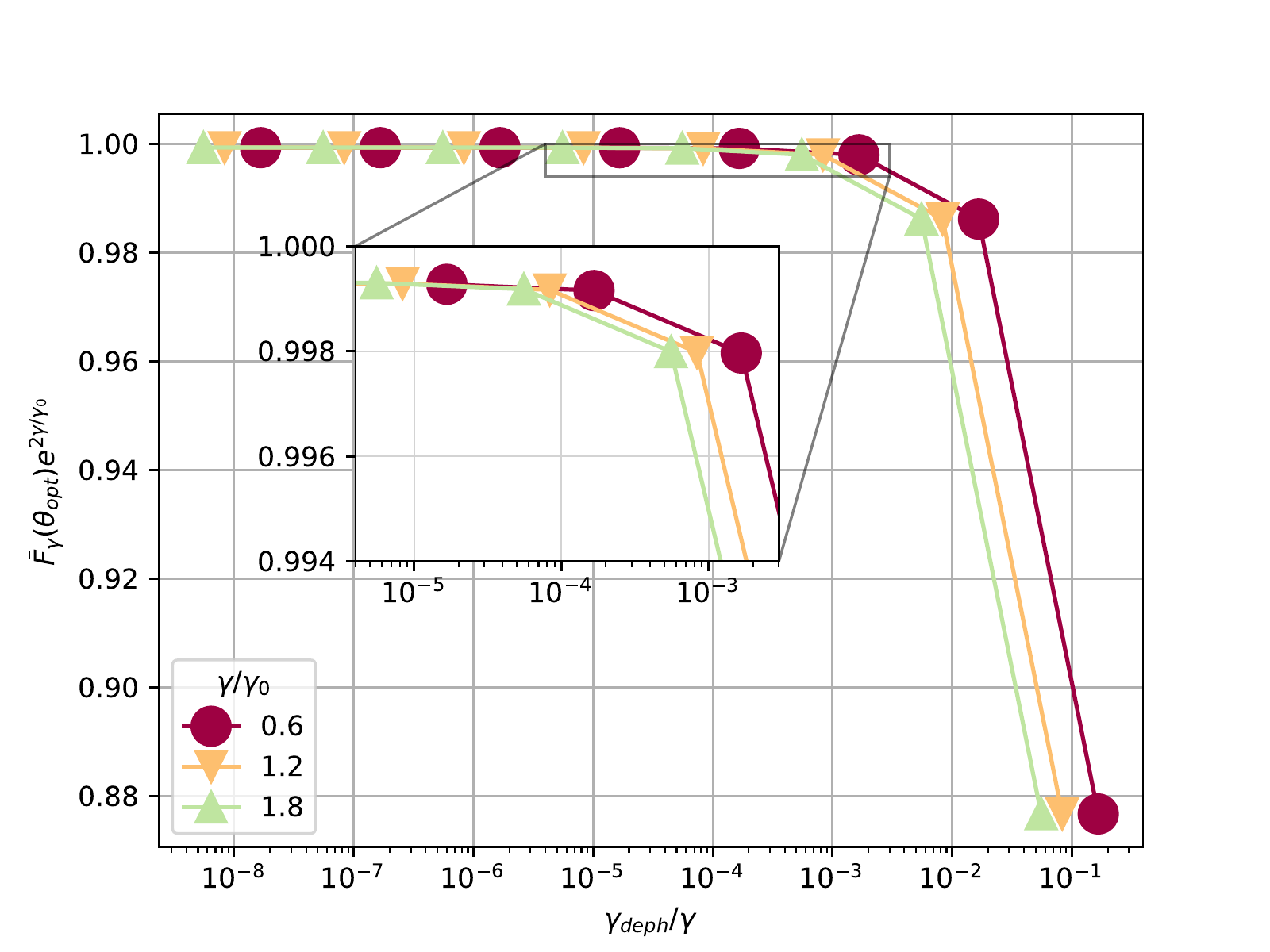}
    \caption{Behavior of the rescaled average gate fidelity on as a function of the pure dephasing rate, $\gamma_{deph}/\gamma_0$, for the quantum circuit having the best average gate fidelity in the unitary case, i.e., corresponding to $U/J_{\mathrm{max}}=0.5$ and 20 blocks. The different set of points correspond to alternative scenarions in which a different particle loss rate, $\gamma/\gamma_0$, is assumed (see the legend). 
    The inset shows a close up on the region where the fidelity starts to decrease, e.g., around $\gamma_{deph}/\gamma_0\sim 10^{-3}$.
    }
    \label{fig:loss_deph_fid}
\end{figure}
 \subsection{Pure dephasing in the presence of particle loss}
 In order to account for other sources of decoherence, we now consider incoherent effects on the fidelity related to pure dephasing noise. In the context of open quantum systems in the Lindblad master equation formalism~\cite{wallsQuantumOptics}, these dissipation terms can be taken into account by means of a Lindblad operator proportional to the photon number in each propagation channel, $n_j=a^{\dagger}_j a_j$, i.e.,  
 \begin{equation}\label{eq:dissipator_pure_dephasing}
     D_{\mathrm{deph}}[\,\rho]=\gamma_{\mathrm{deph}}\sum_{j}\left[n_j \rho\, n_j-\frac{1}{2}\left(n^{2}_j \rho + \rho\, n^{2}_j \right)\right] \, ,
 \end{equation}
 in which $\gamma_{\mathrm{deph}}$ corresponds to the pure dephasing rate (hereafter assumed to be the same in each channel).\\
 Some numerical results concerning the circuit response in the case of the CNOT gate are shown in Fig.~\ref{fig:loss_deph_fid}. There, we report the behavior of the average gate fidelity as a function of the dephasing rate $\gamma_{\mathrm{deph}}/\gamma_0$, rescaled by the factor $e^{2\gamma/\gamma_0}$, which accounts for the single-particle loss associated to population decay (as just detailed in Sec.~\ref{sec:particle_loss_VS_thermal}). In particular, numerical results are obtained by means of the protocol discussed in the previous section, and describe the interplay between particle loss and noise due to pure dephasing on the structure leading to the best average gate fidelity in the unitary regime, i.e. for $\gamma=\gamma_{\mathrm{deph}}=0$. \\
 It is worth noting that realistic pure dephasing rates might strongly depend on the specific material platform employed to encode these photonic qubits. Working with passive photonic integrated circuits (e.g., in silicon or SiN) at low temperature and at visible/telecom wavelengths would ensure negligible contribution to the dephasing of single propagating photons. On the other hand, alternative platforms based, e.g., on exciton polaritons might suffer from phonon-induced dephasing of the excitonic component. For single quantum well excitons, low temperature coupling to the phonon bath as well as spectral diffusion (due, e.g., to small resonance fluctuations due to inhomogeneities in the quantum well thickness) lead to pure dephasing rates that have been measured in the 0.1 $\mu$eV/exciton range [see, e.g., Rohan Singh, PhD thesis, University of Colorado Boulder, 2015]. We further notice that these effects are expected to be reduced by the strong light-matter coupling of exciton and photon fields giving rise to propagating polaritons. Also, considering that an excitonic fraction of 20\%-30\% can be employed to have a reasonable compromise between nonlinearity (due to the excitonic component) and polariton speed (i.e., a group velocity mainly determined by the photonic component), we may expect a polariton pure dephasing rate in the order of $\gamma_{\mathrm{deph}}<0.01\gamma\simeq 0.001\gamma_0$, which according to the results shown in Fig.~\ref{fig:loss_deph_fid} gives us confidence that such decoherence mechanism should not appreciably affect the performance of the optimal two-qubit quantum gates proposed in this work.
 
}
\section{Tolerance to fabrication defects: static parameters fluctuations}
\label{appendix:tolerance}
{In this section we provide further details about the robustness of the theoretical results reported in the manuscript against another possible source of error, namely static noise due, e.g., to parameters fluctuations of the model, as the ones that might be induced, e.g., from fabrication imperfections or non-ideal realization of the designed structures. Indeed, while we believe that the optimization of a multi-port quantum photonic interferometer displaying an almost ideal deterministic two-qubit gate for QC is already an interesting result, it would be of little practical use if not tolerant to static parameters fluctuations with respect to the optimal scenario. Here, we perform a sample test on our optimal structure to show its robustness to different sources of static noise. We restrict the discussion of this statistical analysis to the CNOT operation, with straightforward generalization to the MS gate. The main results are summarized in Fig.~\ref{fig:robustness plot} for the circuit performing the best CNOT discussed in the manuscript. In particular, we consider perturbations on the model structure that gives the results reported in Fig.~2c of the main text.}
\begin{figure*}[t]
    \centering
    \includegraphics[width=0.4\textwidth]{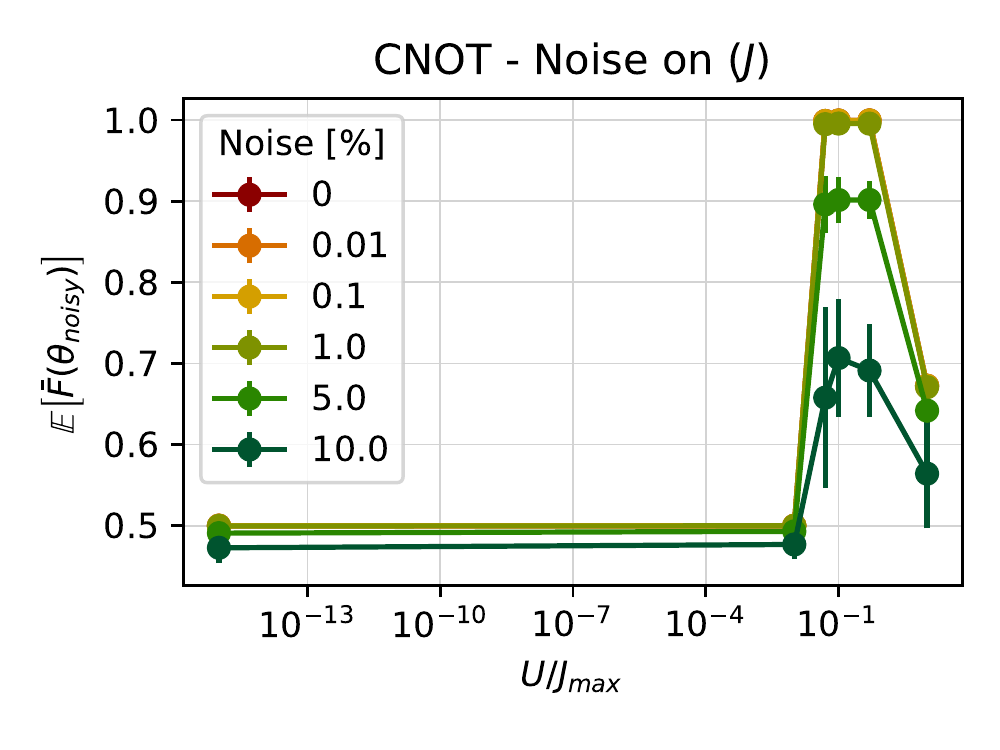}
    \shiftleft{7.5cm}{\raisebox{5.cm}[0cm][0cm]{(a)}}
    \hspace{1cm}
    \includegraphics[width=0.4\textwidth]{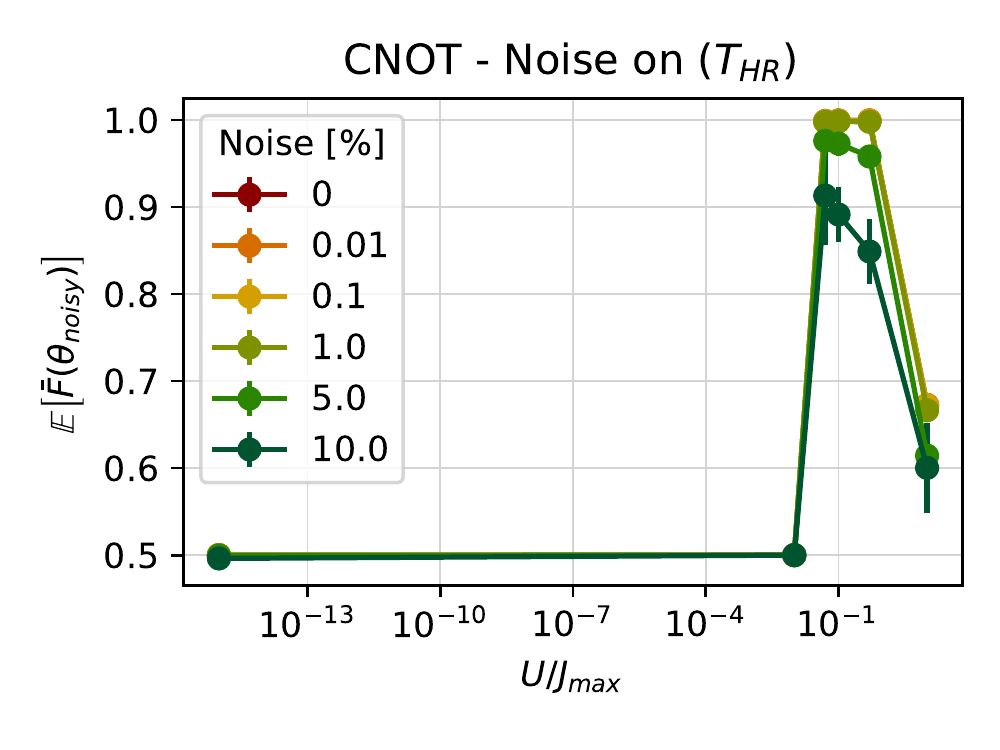}
    \shiftleft{7.5cm}{\raisebox{5.cm}[0cm][0cm]{(b)}}
    \includegraphics[width=0.4\textwidth]{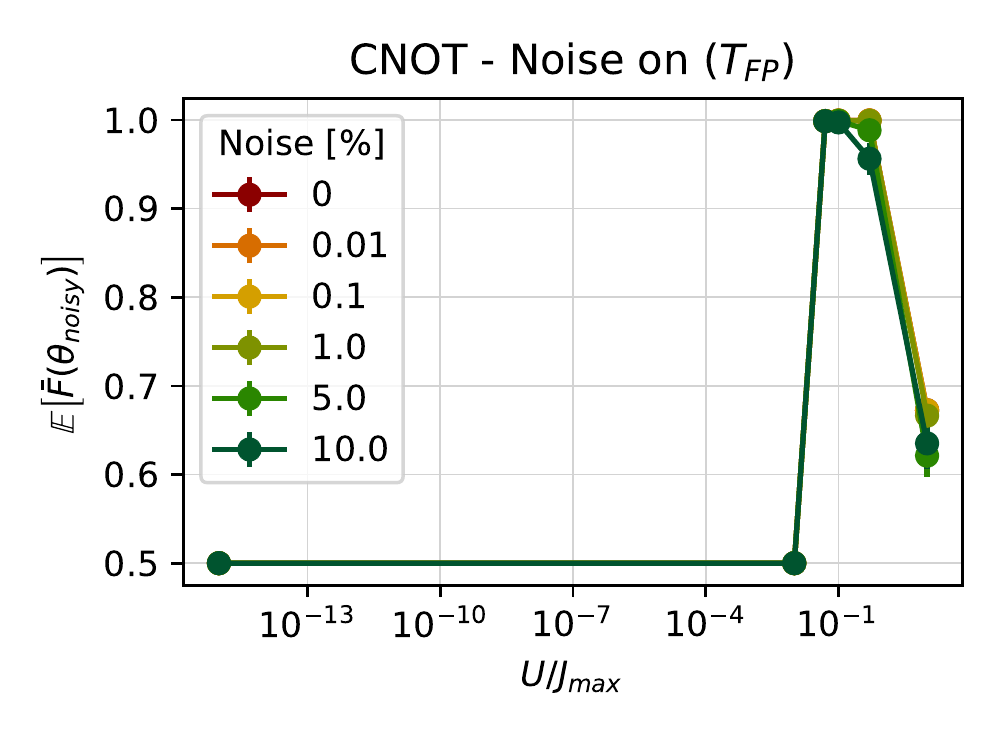}
    \shiftleft{7.5cm}{\raisebox{5.cm}[0cm][0cm]{(c)}}
    \hspace{1cm}
    \includegraphics[width=0.4\textwidth]{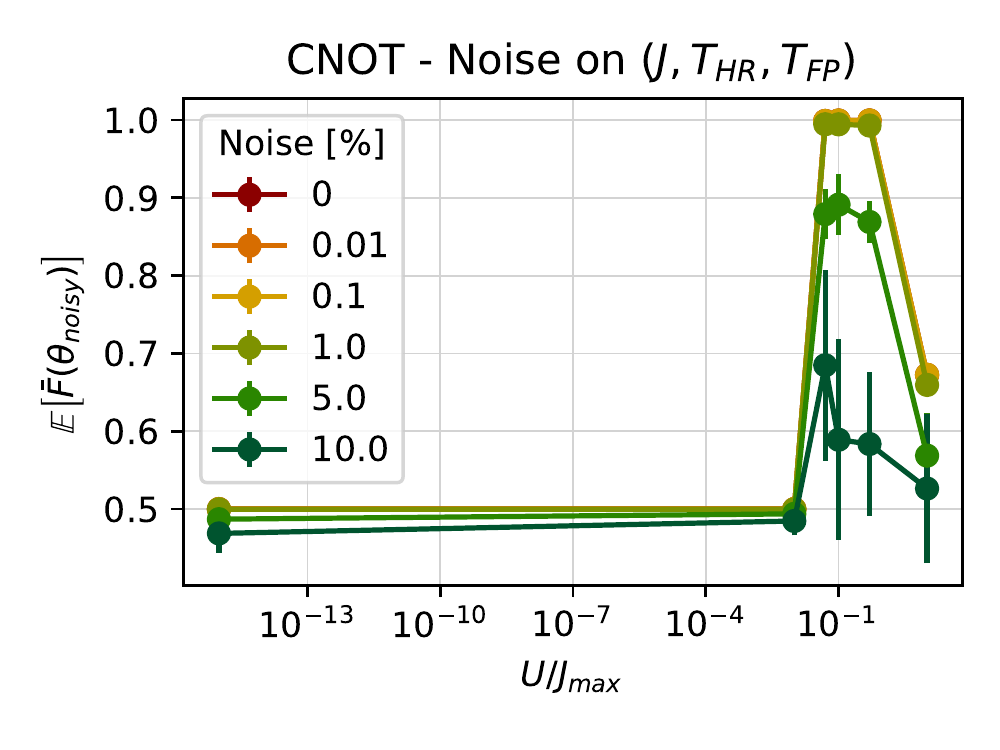}
    \shiftleft{7.5cm}{\raisebox{5.cm}[0cm][0cm]{(d)}}
   \caption{{CNOT tolerance to static parameters fluctuations}, showing the comparison of the average gate fidelity when applying increasing noise levels ($\mathcal{N}_{max}$) to the optimal parameters of the best block structure obtained for different nonlinearity values. The gate performance is tested against \textbf{(a)} variations of the hopping parameters ($J$), \textbf{(b)} the time of interaction ($T_{HR}$), and \textbf{(c)} the free propagation time ($T_{FP}$). Finally, all the previous model parameters are simultaneously perturbed in \textbf{(d)}, showing the model resilience to all sources of noise. All the plots show the average trend upon 20 different sampled error distributions; error bars scale as the standard deviation $\sigma$.}
    \label{fig:robustness plot}
\end{figure*}


{Due to possible fabrication imperfections, the set of parameter values $\{\theta^{*}_{s}\}$  characterizing the fabricated interferometer will deviate from the optimal ones $\{\theta^{opt}_{s}\}$. In addition, since fabrication might also affect the system geometry, even the free propagation and interaction times, i.e., $T_{FP}$ and $T_{HR}$ in our definition, might deviate from the nominal values optimized through the minimization procedure. In what follows, we are going to consider different scenarios. However, for the sake of simplicity, we assume that fabrication leads to uniformly distributed imperfections with respect} {to a fraction $\mathcal{N_{\mathrm{max}}}$ of the reference values $\theta_{ref}\in \{J_{\mathrm{max}}, T_{FP},T_{IR}\}$ used in our discussion, and we study the resilience of the structure under different noise conditions, as shown in Fig.~\ref{fig:robustness plot}. In particular, we define the action of the noise on a set of parameters, $\theta_s$, as follows
\begin{equation}
    \label{eq:noise}
    \theta_s^{noisy}=\theta_s+\mathcal{N}\theta_{ref} \, ,
\end{equation}
where $\mathcal{N}$ is sampled from a uniform distribution in $ \{-\mathcal{N_{\mathrm{max}}}, \mathcal{N_{\mathrm{max}}\}}$. Results describing the effects of fabrication noise on the hopping rates, $T_{HR}$, and  $T_{FP}$ are reported in Fig.~\ref{fig:robustness plot}a,b,c, respectively. Finally, in Fig.~\ref{fig:robustness plot}d we show the average fidelity when simultaneously considering all the sources of static noise separately considered before. All the plots show the average trend upon 20 different sampled error distributions, where error bars are quantified as a single root mean square, $\sigma$ (notice that these are in general visible only for $\mathcal{N}_{max}\ge 5 \% $). In all the possible scenarios, an appreciable discrepancy from the ideal case ($N=0 \%$) is only visible for noise levels above $1\%$ in each perturbed variable, which should be considered significant deviations from the expected outcomes in most platform realizations. Therefore, we can conclude that the optimized configuration is robust, at least against these types of fabrication imperfections. Similar conclusions can be drawn for the MS gate, as shown in Fig~\ref{fig:robustness plot m-s}. }


\begin{figure*}
    \centering
    \includegraphics[width=0.4\textwidth]{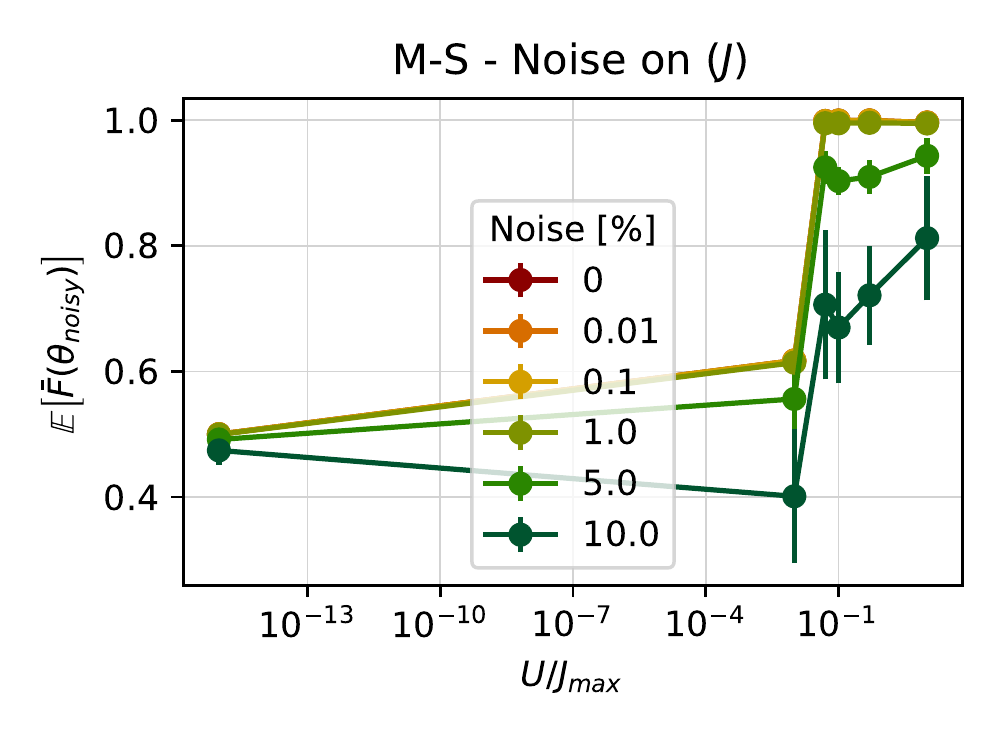}
    \shiftleft{7.5cm}{\raisebox{5cm}[0cm][0cm]{(a)}}
    \hspace{1cm}
    \includegraphics[width=0.4\textwidth]{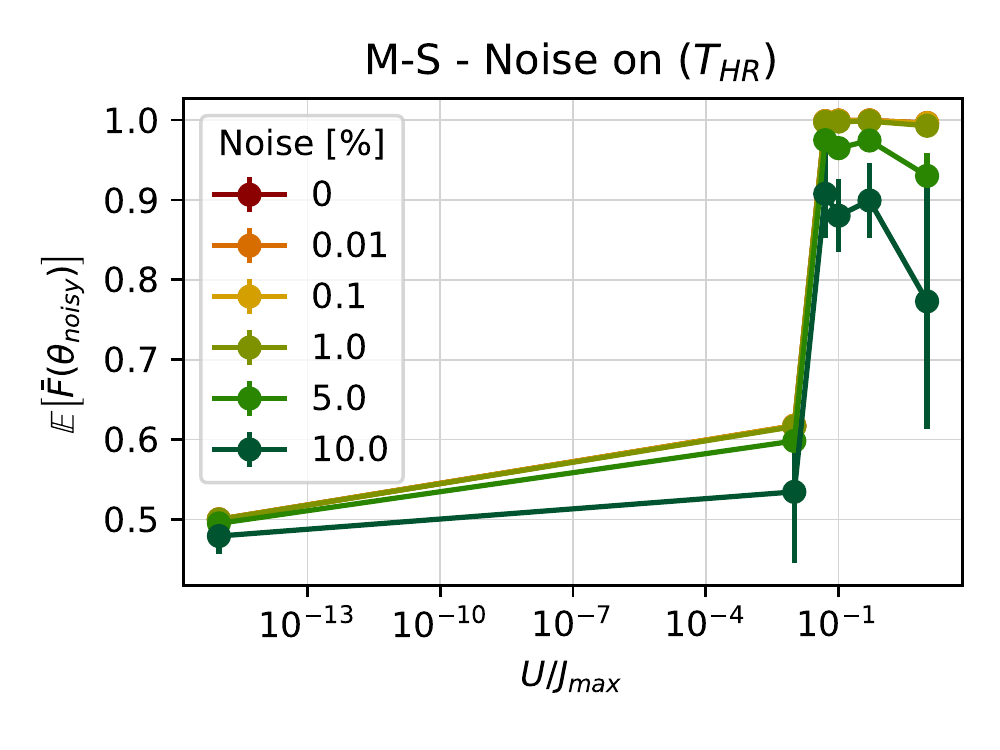}
    \shiftleft{7.5cm}{\raisebox{5cm}[0cm][0cm]{(b)}}
    \includegraphics[width=0.4\textwidth]{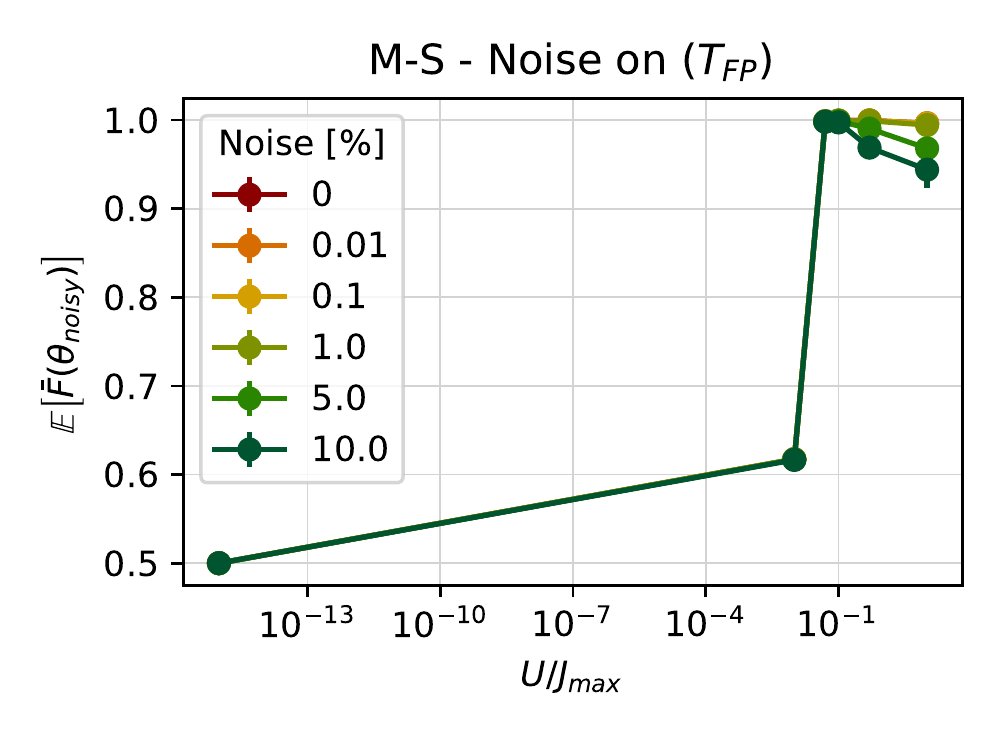}
    \shiftleft{7.5cm}{\raisebox{5cm}[0cm][0cm]{(c)}}
    \hspace{1cm}
    \includegraphics[width=0.4\textwidth]{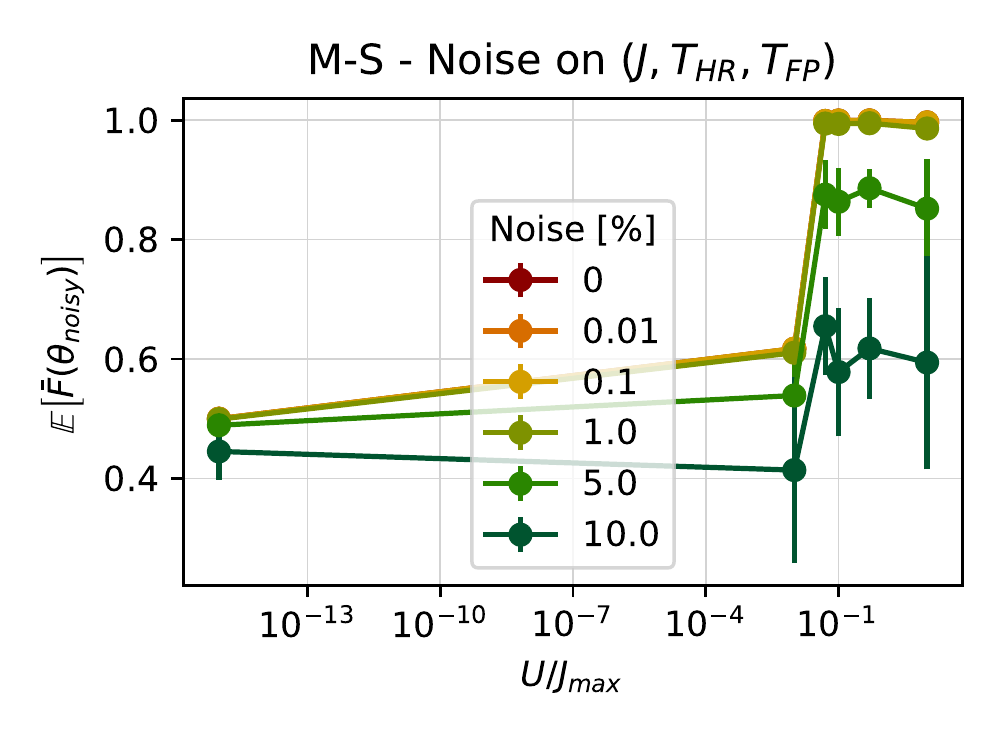}
    \shiftleft{7.5cm}{\raisebox{5cm}[0cm][0cm]{(d)}}
   \caption{{Tolerance of the approximated M-S gate to parameters fluctuations}, obtained from the comparison of the average gate fidelity when applying increasing noise ($\mathcal{N}_{max}$) to the optimal parameters of the optimized block structure, for each different nonlinearity value. The structure is very robust with respect to variations of the hopping parameters ($J$) (a), the time of interaction ($T_{HR}$) (b) and the free propagation time ($T_{FP}$) (c). In (d) we plot the fidelity upon simultaneous perturbation of all the parameters, demonstrating that the model has strong resilience even in the case combined of noise. All the plots show the average trend upon 20 different sampled error distributions, error bars scale as the standard deviation $\sigma$.}
    \label{fig:robustness plot m-s}
\end{figure*}
\section{Extra plots for CNOT and M-S gates}
\label{appendix:extra}

For the sake of completeness, we hereby present additional results complementing the ones presented in the Results section of the main text. In particular, in Fig.~\ref{suppl fig:extra fig 1} we show the optimal M-S full matrix, corresponding to the one displayed in the main text only restricted to the computational space. Even in this case since no logic state is mapped outside the logic space we can conclude that such a gate is deterministic.

\begin{figure*}
    \includegraphics[trim={0 3.5cm 0 .cm},clip,width=0.9\textwidth]{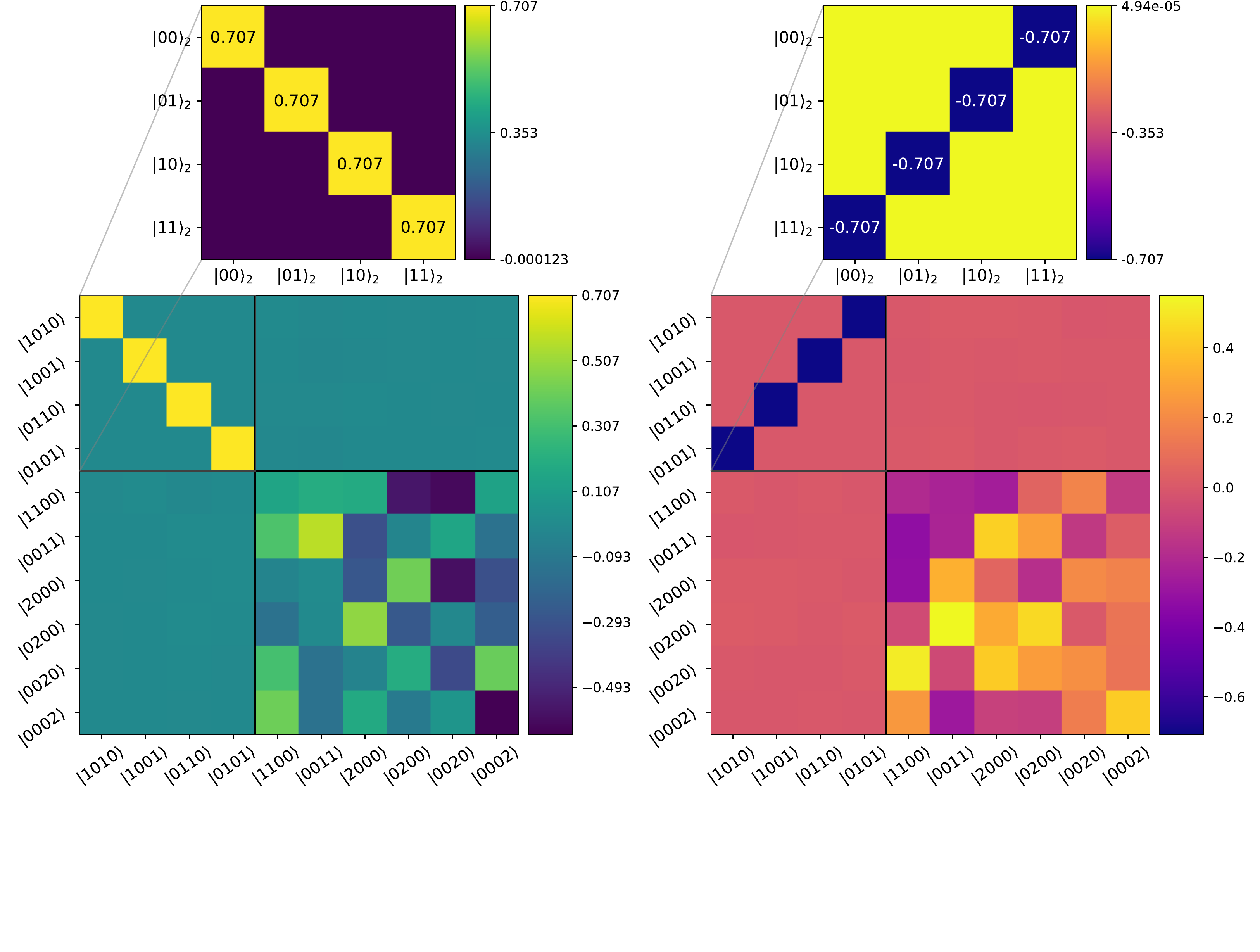}
    \caption{\textbf{Best M-S matrix} Real and imaginary parts of the approximate M-S matrix with all the possible two-photon input/output states, when using a 18 blocks structure with $U=0.5J_{\mathrm{max}}$, for which we get $\bar{F}(\theta_{opt})\approx 99.95\%$. Vertical and horizontal black lines divide the logic space from the one based on states that are not in the computational basis. The projection of logic states off the computational space is negligible, as clearly seen from the plot. The zoom highlights the gate matrix in the two-qubit logic space.}
    \label{suppl fig:extra fig 1}
\end{figure*}


\section{Optimized circuit parameters for CNOT and M-S gates}
\label{appendix:opt params}
In this section, we are reporting tables with the entries of the optimized parameters for the best CNOT and M-S structures, respectively.
\begin{table}
    \centering
    \input{cnot_opt_Js_0.5}
    \caption{Optimal hopping parameters for CNOT structure with 20 blocks and nonlinearity $U=0.5J_{\mathrm{max}}$.}
    \label{tab:cnot J}
\end{table}
\begin{table}
    \centering
    \input{m-s_opt_Js_0.5}
     \caption{Optimal hopping parameters for M-S structure with 18 blocks and nonlinearity $U=0.5J_{\mathrm{max}}$.}
    \label{tab:m-s J}
\end{table}

\end{document}

%% file: cnot_opt_Js_0.5.tex
\begin{tabular}{lrrrrr}
\toprule
{} &   $J_{0}/J_{\mathrm{max}}$ &   $J_{1}/J_{\mathrm{max}}$ &   $J_{2}/J_{\mathrm{max}}$ &   $J_{3}/J_{\mathrm{max}}$ &  $J_{4}/J_{\mathrm{max}}$ \\
\midrule
1 & 0.397053 & 0.460288 & 0.433919 & 0.345798 & 0.349151 \\
2 & 0.357316 & 0.695590 & 0.589448 & 0.738313 & 0.501163 \\
3 & 0.149630 & 0.624371 & 0.614882 & 0.261721 & 0.313165 \\
4 & 0.311026 & 0.159906 & 0.441268 & 0.446604 & 0.766868 \\
5 & 0.305264 & 0.164163 & 0.656376 & 0.135338 & 0.497878 \\
6 & 0.502554 & 0.527935 & 0.347979 & 0.794379 & 0.595067 \\
7 & 0.284354 & 0.463411 & 0.526538 & 0.647302 & 0.224454 \\
8 & 0.603812 & 0.425294 & 0.307670 & 0.413925 & 0.673114 \\
9 & 0.806110 & 0.437601 & 0.716100 & 0.607492 & 0.602587 \\
10 & 0.220781 & 0.404325 & 0.765186 & 0.482268 & 0.655072 \\
11 & 0.053196 & 0.585119 & 0.286334 & 0.570954 & 0.402904 \\
12 & 0.631146 & 0.603053 & 0.175298 & 0.493723 & 0.995993 \\
13 & 0.620809 & 0.365945 & 0.729352 & 0.540424 & 0.771402 \\
14 & 0.422265 & 0.480154 & 0.346867 & 0.523882 & 0.665364 \\
15 & 0.820865 & 0.360282 & 0.635051 & 0.769726 & 0.538203 \\
16 & 0.383178 & 0.371631 & 0.443623 & 0.495927 & 0.472496 \\
17 & 0.493759 & 0.891368 & 0.449298 & 0.681804 & 0.448260 \\
18 & 0.645425 & 0.005713 & 0.081965 & 0.191282 & 0.581207 \\
19 & 0.389986 & 0.316030 & 0.898467 & 0.188049 & 0.769573 \\
20 & 0.414400 & 0.445733 & 0.662107 & 0.421034 & 0.576480 \\
\bottomrule
\end{tabular}

%% file: m-s_opt_Js_0.5.tex
\begin{tabular}{lrrrrr}
\toprule
{} &   $J_{0}/J_{\mathrm{max}}$ &   $J_{1}/J_{\mathrm{max}}$ &   $J_{2}/J_{\mathrm{max}}$ &   $J_{3}/J_{\mathrm{max}}$ &  $J_{4}/J_{\mathrm{max}}$ \\
\midrule
1 & 0.202968 & 0.335635 & 0.283428 & 0.260127 & 0.727614 \\
2 & 0.275506 & 0.633033 & 0.414959 & 0.289816 & 0.605048 \\
3 & 0.274936 & 0.402640 & 0.453810 & 0.096417 & 0.206710 \\
4 & 0.571623 & 0.305405 & 0.515797 & 0.182360 & 0.617496 \\
5 & 0.464091 & 0.424111 & 0.552958 & 0.544361 & 0.394582 \\
6 & 0.311228 & 0.576185 & 0.633629 & 0.556476 & 0.505165 \\
7 & 0.526125 & 0.452571 & 0.729350 & 0.403067 & 0.640229 \\
8 & 0.367590 & 0.499176 & 0.500112 & 0.267933 & 0.153150 \\
9 & 0.700993 & 0.228619 & 0.546548 & 0.315510 & 0.186718 \\
10 & 0.104315 & 0.146993 & 0.275102 & 0.604278 & 0.381363 \\
11 & 0.464906 & 0.587760 & 0.661208 & 0.344834 & 0.557134 \\
12 & 0.485208 & 0.219439 & 0.513613 & 0.419543 & 0.263057 \\
13 & 0.340085 & 0.197179 & 0.470652 & 0.501840 & 0.464358 \\
14 & 0.053501 & 0.181264 & 0.682423 & 0.395839 & 0.408464 \\
15 & 0.535356 & 0.185371 & 0.925122 & 0.004042 & 0.718715 \\
16 & 0.180504 & 0.403074 & 0.747207 & 0.420334 & 0.209329 \\
17 & 0.327039 & 0.360125 & 0.410214 & 0.375598 & 0.310273 \\
18 & 0.060446 & 0.002149 & 0.286505 & 0.446174 & 0.086044 \\
\bottomrule
\end{tabular}